\documentclass[12pt,a4paper]{article}
\usepackage{amsmath}
\usepackage{amssymb,latexsym}
\usepackage{natbib}
\usepackage{graphicx}
\usepackage{float}
\usepackage{xcolor}
\usepackage{url}
\usepackage{multirow}
\usepackage{enumitem}

\newcommand {\bx}{{\bf x}}
\newcommand {\bX}{{\bf X}}
\newcommand {\bu}{{\bf u}}
\newcommand {\bU}{{\bf U}}
\newcommand {\bv}{\boldsymbol {v}}
\newcommand {\bA}{{\bf A}}
\newcommand {\ba}{{\bf a}}
\newcommand {\bB}{{\bf B}}

\newcommand {\bR}{{\bf R}}
\newcommand {\bzero}{{\bf 0}}

\def\thick#1{\hbox{\rlap{$#1$}\kern0.4pt\rlap{$#1$}\kern0.25pt$#1$}}

\title{\vspace{-1cm}Examining Collinearities}
\author{\textsc{Zillur R. Shabuz}$^{a}$ \& \textsc{Paul H. Garthwaite}$^{b}$\\
\normalsize $^{a}$Department of Statistics, University of Dhaka, Dhaka, Bangladesh\\
\normalsize $^{b}$Department of Mathematics and Statistics, The Open University, Milton Keynes, UK}
\date{}

\begin{document}

\maketitle

\
\

\
\

\textbf{Corresponding author} \\ \textsc{Zillur R. Shabuz} \\ Associate Professor \\ Department of Statistics \\ University of Dhaka \\ Bangladesh \\ e-mail: zrshabuz@du.ac.bd 

\newpage

\begin{center}
\LARGE{Examining Collinearities}
\end{center}

\begin{abstract}
The cos-max method is a little-known method of identifying collinearities. It is based on the cos-max transformation [4], which makes minimal adjustment to a set of vectors to create orthogonal components with a one-to-one correspondence between the original vectors and the components. The aim of the transformation is that each vector should be close to the orthogonal component with which it is paired. Vectors involved in a collinearity must be adjusted substantially in order to create orthogonal components, while other vectors will typically be adjusted far less. The cos-max method uses the size of adjustments to identify collinearities. It gives a coherent relationship between collinear sets of variables and variance inflation factors (VIFs), and identifies collinear sets using more information than traditional methods. In this paper we describe these features of the method and examine its performance in examples, comparing it with alternative methods. In each example the collinearities identified by the cos-max method only contained variables with high VIFs and contained all variables with high VIFs. The collinearities identified by other methods did not have such a close link to VIFs. Also, the collinearities identified by the cos-max method were as simple or simpler than those given by other methods, with less overlap between collinearities in the variables that they contained.
\end{abstract}

\noindent
Keywords: {Auxiliary regression; cos-max; eigenvector analysis; variance inflation factor; variance decomposition}

\subsection*{1. Introduction} \label{S:Introduction}

When regressors are orthogonal, the use and interpretation of a multiple regression model is straightforward but, in practice, it is quite common for some of the regressors to have high correlation. More generally, there may be a near-linear relationship amongst the regressors, in which case the variables are said to be collinear or, if there is a perfect linear relationship between some of them, there is an exact collinearity. A dataset may exhibit more than one collinearity and then identifying the relationships that exist between the regressors can be challenging. Two relatively new methods for examining collinearities are proposed in Garthwaite et al. [4]. Collinearities form a very small part of that paper and the methods have attracted little attention. The two methods are called the cos-max method and the cos-square method. They are quite similar and here we restrict attention to the cos-max method. Through examples we show that it is informative and can be a more informative tool for examining collinearities than traditional methods.

Collinearity has a number of practical consequences. When there is an exact collinearity, the ordinary least squares (OLS) estimate of $\boldsymbol{\beta}$ is not unique and variances of some individual estimates are infinite [5]. When there is approximate collinearity, OLS estimates have large variances and covariances, and one or more important regressors may have low $t$-ratios and overly wide confidence intervals, even if the overall model parameters are significant and the model has a high ${R}^2$ value. Also, OLS estimates and their variances will be sensitive to minor changes in the data [5] and the deletion of observations from the data or the addition of new observations can change the regressors that variable selection chooses to include in the model. Collinearity also tends to inflate some of the OLS estimates in absolute value and, as a result, the average squared distance between parameters and their estimates can become large [11].  In some cases one or more of the estimates may even produce incorrect signs of the OLS estimates, contradicting the relationship between these regressors and the response variable [6].   

The cos-max method is described in Section 2 and two other methods of examining collinearity are briefly described in Section 3. These latter methods are eigenvector analysis and variance decomposition. The focus of Section 3 is the application of these methods and the cos-max method to four examples in which there are collinearities. The third example is the most complex; six of its regressors are broadly contained in a four-dimensional subspace and using two collinearities to describe the relationships between the variables may be overly simplistic. The first three examples use real data while the fourth example uses artificial data in which there are overlapping collinearities. An advantage from constructing artificial data is that the true structure underlying the data is known. Concluding comments are given in Section 4.

\subsection*{2. The cos-max method} \label{S:CosMax}

The cos-max method is based on the  cos-max transformation [4], which makes minimal adjustment to a set of vectors $\bx _1, \ldots, \bx_m$ so as to form a set of orthonormal components $\bu_1, \ldots, \bu_m$. Principal components is the best known transformation that yields orthogonal components but, unlike principal components, the cos-max transformation aims to retain the identity of the $x$ variables; $\bx_i$ is associated with $\bu_i$. Specifically, the cos-max transformation determines vectors $ \bu_1, \ldots, \bu_m$ such that
\begin{equation} \label{eqPsi}
\psi = \sum_{i=1}^m \bx_i^T \bu_i
\end{equation}
is maximised, subject to $ \bu_i^T \bu_j =0$ for $i \neq j$ and $ \bu_i^T \bu_i =1$ for each $i$. Put $\bX = (\bx_1, \ldots, \bx_m)$ and $\bU = (\bu_1, \ldots, \bu_m)$. Garthwaite et al. [4] (Corollary 1) show that
\begin{equation} \label{EqTran}
\bU = \bX \bA
\end{equation}
where $\bA$ is the symmetric positive-definite square root of $(\bX^T \bX) ^{-1}$. The matrix $\bA$ is unique and is referred to as the transformation matrix.

We suppose $\bx_1, \ldots, \bx_m$ are $n \times 1$ vectors of observations on regressors $X_1,\ldots, X_m$. Throughout the remainder of this paper we also suppose that the data have been standardised so that, for each $i$,  $ \bx_i^T \bx_i =1$  and the elements of $\bx_i$ sum to 0. Then $\bx_i^T \bu_i$ is the correlation between $\bx_i$ and $\bu_i$ and, from (1), the cos-max transformation maximises the sum of these correlations. Also, if $\bx_i$ and $\bu_i$ denote directional vectors, then $\bx_i^T \bu_i$ is the cosine of the angle between them, which gives the cos-max transformation its name.

The cos-max method judges which variables are involved in collinearities by examining  $\bA$, the symmetric square-root of the inverse of the correlation matrix [i.e., $\bA\bA = (\bX^T \bX)^{-1}$ ].  In the context of collinearity, $\bA$ has two crucial features.
\begin{enumerate}
\item Let $\ba_i$ denote the $i$th column of $\bA$ and let $a_{ji}$ denote the $j$th element of $\ba_i$. Then equation (\ref{EqTran}) gives
\begin{equation} \label{EqTran2}
\bu_i = \bX \ba_i  = \sum_{j=1}^m a_{ji}\bx_j .
\end{equation}
We will refer to $\bu_i$ as the {\em surrogate} of $\bx_i$. If $a_{ji}$ is close to 0, then $\bx_j$ has little influence on the relationship between $\bx_i$ and its surrogate, while if $|a_{ji}|$ is large, then $\bx_j$ has a clear influence on the relationship between $\bx_i$ and its surrogate. Obviously $a_{ii}$ should be large so that $\bx_i$ is highly correlated with its surrogate. Which other elements of $\bA$ will be large and which will be small is discussed more fully in an appendix. Essentially, a large value for $a_{ji}$ (which equals $a_{ij}$ as $\bA$ is a symmetric matrix) indicates an association between $\bx_i$ and $\bx_j$, while a small value suggests little association. 
\item Variance inflation factors (VIFs) are the most commonly used method of identifying the existence of collinearities and which variables are involved in collinearities.  Let $R_i^2$ be the coefficient of determination when $X_i$ is regressed on the remaining $m-1$ regressors. That is,   $R_i^2$ is the proportion of the variation in $X_i$ that can be predicted by the other $m-1$ regressors. The VIF of $X_i$ is defined to be
\begin{equation} \label{EqVIF}
\mbox{VIF}_i = (1-R_i^2)^{-1},
\end{equation}
so a large value of VIF$_i$ shows that $X_i$ is involved in a collinearity. It is well-known that VIF$_i$ is equal to the $i$th diagonal element of $(\bX^T \bX)^{-1}$. Hence, if $\widehat{\beta}_i$ is the $i$th estimated regression coefficient in a regression model with regressors $X_1, \ldots, X_m$ and random error variance  $\sigma^2 $, then var$(\widehat{\beta}_i) =\sigma^2 \mbox{VIF}_i$. Thus VIF$_i$ is the factor by which the variance of $\widehat{\beta}_i$ is inflated due to linear dependencies among the regressors.

There is no theory-based rule for what values of a VIF should be equated to collinearity, but common rules of thumb are that VIFs above 5 or 10 indicate marked collinearity [9,10]. 
As VIF$_i$ is equal to the $i$th diagonal element of $(\bX^T \bX)^{-1}$ and $\bA$ is the symmetric square-root of $(\bX^T \bX)^{-1}$, it follows that
\begin{equation} \label{EqVIF2}
\mbox{VIF}_i = \ba_i^T \ba_i .
\end{equation}
This is the second crucial feature of the $\bA$ matrix.
\end{enumerate}

To identify variables involved in collinearities, the cos-max method determines $\bA = (\ba_1, \ldots, \ba_m)$ from $\bX^T \bX$ and calculates $\ba_1^T \ba_1, \ldots, \ba_m^T \ba_m$. From point 2 (equation (\ref{EqVIF2})), the later are the VIFs of $X_1, \ldots, X_m$. If $X_i$ were uncorrelated with the other $X$ variables, then the $i$th element of $\ba_i$ would be 1, the other elements of $\ba_i$ would be 0, and the VIF of $X_i$ would be 1. If $X_i$ has a large VIF (above 5, say), then some elements of $\ba_i$ must be large. From point 1, the elements of $\ba_i$ that are large correspond to those $X$ variables that have an association with $X_i$. The cos-max method compares the elements of $\ba_i$ with some threshold (in our examples $0.75$ is used as the threshold) and $X_j$ is  deemed to be involved in the collinearity of $X_i$ if the $j$th element of $\ba_i$ exceeds the threshold.

This is a coherent approach to determining whether $X_i$ has a large VIF and, if it is large, identifying which variables cause it to be large: performing each of these functions is based on $\ba_i$. An attractive feature of the cos-max method is that it uses more information to examine collinearities than standard alternative methods. The latter typically examine the eigenvectors that are associated with small eigenvalues, while the cos-max method examines one of $\ba_1, \ldots, \ba_m$ for each VIF that is large.  Eigenvectors are the same size as the $\ba_i$ (they are $m \times 1$ vectors) but the number of large VIFs will always exceed the number of small eigenvalues.

\subsection*{3. Examples} \label{S:Exam}

\subsubsection*{3.1 Example 1: Sales Data (one collinearity)} \label{S:Ex1}
Data reported in Chatterjee and Hadi [3] were collected by a firm over a period of 23 years. The firm had fairly stable operating conditions during the period of data collection. Their objective was to regress aggregate sales against five regressors: advertising expenditure $(X_1)$, promotion expenditure $(X_2)$, sales expense $(X_3)$, advertising expenditure in the previous year $(X_4)$, and lagged promotion expenditure $(X_5)$. Table \ref{corsales} displays the correlations among the regressors and their VIFs. The correlations are small but four of the regressors have high VIFs ($X_1$, $X_2$, $X_4$ and $X_5$). Here we examine ways of identifying collinearities among the regressors.

\begin{linespread}{1.5}
\begin{table}
\caption{Correlation matrix  and VIFs of regressors for the sales data} \label{corsales} \vspace{-0.05in}
\newcommand\Fontvi{\fontsize{10.0}{10.0}\selectfont}
\Fontvi
\begin{center}
\begin{tabular}{ l  c c c c c c}
\hline \vspace{-0.2in} \\ 
Variable  & $X_1$ & $X_2$ & $X_3$ & $X_4$ & $X_5$ & \ \ \ VIF \\
\hline \vspace{-0.15in} \\
$X_1$ &  \ 1.000 & -0.357 & -0.129 & -0.140 & -0.496 & \ \ \  36.94\\
$X_2$ &  -0.357 & \ 1.000 & \ 0.063 & -0.316 & -0.296 & \ \ \  33.47\\
$X_3$ &  -0.129 & \ 0.063 & \ 1.000 & -0.166 & \ 0.208 & \ \ \ \ 1.08 \\
$X_4$ &  -0.140 & -0.316 & -0.166 & \ 1.000 & -0.358 & \ \ \  25.92\\
$X_5$ &  -0.496 & -0.296 & \ 0.208 & -0.358 & \ 1.000 & \ \ \  43.52\\
\hline
\vspace{-.4in}
\end{tabular}
\end{center}
\end{table}
\end{linespread} 

Eigenvector analysis is commonly used to examine collinearity. Eigenvalues and eigenvectors of the correlation matrix are determined and near-zero eigenvalues suggest a collinearity. The corresponding eigenvectors are used to identify the $X$-variables involved; in principle, components of the eigenvector that are large in magnitude should correspond to those $X$-variable that are most influencing the collinearity. {\em Condition indices} are sometimes used as the criteria for whether an eigenvalue indicates a collinearity. The condition index of the $j$th  eigenvalue, $\lambda _j$, is defined to be (maximum eigenvalue$)/\lambda_j$.  Standard statistical packages for collinearity diagnostics automatically output condition indices. (Some packages, and some writers, prefer to define condition indices in terms of the singular values of $\bX$, rather than the eigenvalues of $\bX^T\bX$, when the former are the square-roots of the latter.) With the definition used here, a condition index between 100 and 1000 indicates moderate to strong collinearity and a condition index above 1000 is often used to indicate severe collinearity [5,9].

Eigenvalues, eigenvectors and condition indices for the correlation matrix of the sales data (Table \ref{corsales}) are given in Table \ref{eigensales}. The 5th eigenvalue of 0.007 is very small and  the condition index of 233.9 confirms there is a collinearity. None of the other eigenvalues is small so there is only one collinearity in the dataset. Large entries in the 5th eigenvector indicate that the regressors forming the collinearity are $X_1$, $X_2$, $X_4$ and $X_5$. As would be expected, these are the regressors with the large VIFs in Table \ref{corsales}. 

\begin{linespread}{1.5} 
\begin{table}
\caption{Eigenvector analysis and condition indices for the sales data} \label{eigensales} \vspace{-0.05in}
\newcommand\Fontvi{\fontsize{10.0}{10.0}\selectfont}
\Fontvi
\begin{center}
\begin{tabular}{ c c c c c c c c }
\hline \vspace{-0.2in} \\ 
 & \multicolumn{5}{c}{Eigenvector loadings} &   &  Condition \\
\cline{2-6}  \vspace{0.02in}
Eigenvector & $X_1$ & $X_2$ & $X_3$ & $X_4$ & $X_5$ & Eigenvalue & Index \\
\hline \vspace{-0.15in} \\
$\boldsymbol{v}_1$ & \ 0.532 & -0.232 & -0.389 & \ 0.395 & -0.596 & 1.701 & \ \ \ 1.0 \\
$\boldsymbol{v}_2$ & \ 0.024 & -0.825 & \ 0.022 & \ 0.260 & \ 0.501 & 1.288 & \ \ \ 1.3 \\
$\boldsymbol{v}_3$ & \ 0.668 & -0.158 & \ 0.217 & -0.692 & \ 0.057 & 1.145 & \ \ \ 1.5 \\
$\boldsymbol{v}_4$ & -0.074 & \ 0.037 & -0.895 & -0.338 & \ 0.279 & 0.859 & \ \ \ 2.0 \\
$\boldsymbol{v}_5$ & \ \textcolor{blue}{0.514} & \ \textcolor{blue}{0.489}& -0.010 & \ \textcolor{blue}{0.428} & \ \textcolor{blue}{0.559} & 0.007 & 233.9  \\
\hline \vspace{-.4in}
\end{tabular}
\end{center}
\end{table}
\end{linespread}

Variance decomposition is the other well-known method of examining collinearities. The VIF of $X_i$ can be related to eigenvalues and eigenvalues by the equation 
\begin{equation} \label{VarDecom}
\mbox{VIF}_i= \sum_{j=1}^m v_{ij}^2 / \lambda_j
\end{equation}
where $ \lambda_j$ is the $j$th eigenvalue of $\bX^T \bX$ and $ v_{ij}$ is is the $i$th element of the corresponding eigenvector. The variance decomposition method calculates
\begin{equation} \label{PiDef}
\pi _{ji} = \frac{ v_{ij}^2 / \lambda_j}{\mbox{VIF}_i} .
\end{equation}
As var$(\widehat{\beta}_i)  =\sigma^2 \mbox{VIF}_i$, the proportion of var$(\widehat{\beta}_i)$ that can attributed to the $j$th eigenvector is $\pi_{ji}$. This proportion ($\pi_{ji}$) is referred to as the $(j,i)$th variance-decomposition proportion (VDP). A VDP greater than 0.5 that corresponds to a small eigenvalue is a recommended criteria for whether a regressor is part of the collinearity associated with that eigenvalue [2,9].

\begin{linespread}{1.5} 
\begin{table}
\caption{Eigenvalues and variance-decomposition proportions for the sales data} \label{vdpsales} \vspace{-0.05in}
\newcommand\Fontvi{\fontsize{10.0}{10.0}\selectfont}
\Fontvi
\begin{center}
\begin{tabular}{ c c c c c c c c }
\hline \vspace{-0.2in} \\ 
& \multicolumn{5}{c}{Variance-decomposition proportion} \\
\cline{2-6} \vspace{-0.2in} \\
Eigenvalue & $X_1$ & $X_2$ & $X_3$ & $X_4$ & $X_5$ \\
\hline \vspace{-0.15in} \\
1.701 & 0.005 & 0.001 & 0.083 & 0.004 & 0.005 \\
1.288 & 0.000 & 0.016 & 0.000 & 0.002 & 0.004 \\
1.145 & 0.011 & 0.001 & 0.038 & 0.016 & 0.000 \\
0.859 & 0.000 & 0.000 & 0.867 & 0.005 & 0.002 \\
0.007 & \textcolor{blue}{0.985} & \textcolor{blue}{0.983} & 0.012 & \textcolor{blue}{0.973} & \textcolor{blue}{0.989} \\
\hline	\vspace{-0.4in} \\			
\end{tabular}
\end{center}
\end{table}
\end{linespread}

Table \ref{vdpsales} presents the VDPs and eigenvalues for the sales data. As noted earlier, the eigenvalues indicate a single collinearity as the fifth eigenvalue is small (0.007) while other eigenvalues are quite large. The VDPs associated with the fifth eigenvalue are large for $X_1$, $X_2$, $X_4$ and $X_5$ (equalling 0.985, 0.983, 0.973 and 0.989), while $X_3$'s value of 0.012 is well below the threshold of 0.5. Hence variance  decomposition indicates a collinearity between $X_1$, $X_2$, $X_4$ and $X_5$, in agreement with the eigenvector analysis.

To examine collinearity using the cos-max method, the transformation  matrix $\bA=(\bX^T\bX)^{-1}$ was determined from the correlation matrix of the regressors. The columns of $\bA$ are $\ba_1, \ldots.\ba_5$ and their values are given in Table \ref{garsales1}, together with the VIFs. The VIFs may be derived from the transformation matrix. For example, 33.47 is the VIF for $X_2$ and equals $\ba_2^T\ba_2$. As $X_1$, $X_2$, $X_4$ and $X_5$ have large VIFs, we examine $\ba_1$, $\ba_2$, $\ba_4$ and $\ba_5$. The variables $X_1$, $X_2$, $X_4$ and $X_5$ each give a component of $\ba_1^T$ that is large in magnitude, comfortably exceeding a threshold of 0.75, indicating a collinearity between these four variables. The same conclusion is obtained from examining $\ba_2^T$, or $\ba_4^T$, or $\ba_5^T$.

The collinear set identified for this simple example is the same with all methods. However, the eigenvector analysis uses a single row of Table \ref{eigensales} to identify the collinear set, and the variance decomposition method uses a single row of Table \ref{vdpsales}, while the cos-max method more than duplicates the amount of information needed to identify the collinear set. (The cos-max method uses four rows of Table \ref{garsales1}, but the amount of information is not quadrupled as $\bA$ is symmetric). The additional information provided by the cos-max method can add clarity when there is more than one collinearity, as will be seen in the next examples.

\begin{linespread}{1.5} 
\begin{table}
\caption{Cos-max transformation matrix and VIFs for the sales data} \label{garsales1} \vspace{-0.05in}
\newcommand\Fontvi{\fontsize{10.0}{10.0}\selectfont}
\Fontvi
\begin{center}
\begin{tabular}{  c c c c c c c }
\hline \vspace{-0.2in} \\ 
& $X_1$ & $X_2$ & $X_3$ & $X_4$ & $X_5$ & VIF  \\
\hline \vspace{-0.15in} \\
$\boldsymbol{a}_1^{\top}$ & \ \textcolor{blue}{3.743} & \ \textcolor{blue}{2.736} & -0.010 & \textcolor{blue}{2.345} & \ \textcolor{blue}{3.154} & 36.94 \vspace{0.01in} \\ 
$\boldsymbol{a}_2^{\top}$ & \ \textcolor{blue}{2.736} & \ \textcolor{blue}{3.470} & -0.070 & \textcolor{blue}{2.285} & \ \textcolor{blue}{2.952} & 33.47 \vspace{0.01in} \\
$\boldsymbol{a}_3^{\top}$ & -0.010 & -0.070 & \ 1.026 & 0.024 & -0.134 & \ 1.08 \vspace{0.01in} \\
$\boldsymbol{a}_4^{\top}$ & \ \textcolor{blue}{2.345} & \ \textcolor{blue}{2.285} & \ 0.024 & \textcolor{blue}{2.901} & \ \textcolor{blue}{2.604} & 25.92 \vspace{0.01in} \\
$\boldsymbol{a}_5^{\top}$ & \ \textcolor{blue}{3.154} & \ \textcolor{blue}{2.952} & -0.134 & \textcolor{blue}{2.604} & \ \textcolor{blue}{4.249}&43.52\\
\hline \vspace{-.4in}
\end{tabular}
\end{center}
\end{table}
\end{linespread}
 
\subsubsection*{3.2 Example 2: Pitprop Data (three disjoint collinearities)} \label{S:Ex2}

Data used by Jeffers [7] were collected in East Anglia over a period of 10 years to determine the physical characteristics that influence the maximum compressive strength of pitprops made of Corsican pine. The study has 180 pitprops. The physical variables on each pitprop were top diameter ($X_1$), length ($X_2$), moisture content as a percentage of the dry weight ($X_3$), specific gravity at the time of the test ($X_4$), oven-dry specific gravity ($X_5$), number of annual rings at the top ($X_6$), number of annual rings at the base ($X_7$), maximum bow ($X_8$), distance of the point of maximum bow from the top ($X_9$), number of knot whorls ($X_{10}$), length of clear prop from the top ($X_{11}$), average number of knots per whorl ($X_{12}$), and average diameter of the knots ($X_{13}$).
The dataset was used in Garthwaite et al. [4] to illustrate their new methods for identifying collinearities, focusing on the cos-square method and comparing it with eigenvector analysis. Here we focus on the cos-max method and compare it with a broader range of alternatives.

\begin{linespread}{1.5}
\begin{table}
\caption{\footnotesize Correlation matrix for the pitprops data}\label{corpitprop} \vspace{-0.05in}
\newcommand\Fontvi{\fontsize{9.0}{9.0}\selectfont}
\Fontvi
\begin{center}
\begin{tabular}{c@{\hskip 0.25cm} c@{\hskip 0.25cm} c@{\hskip 0.25cm} c@{\hskip 0.25cm} c@{\hskip 0.25cm} c@{\hskip 0.25cm} c@{\hskip 0.25cm} c@{\hskip 0.25cm} c@{\hskip 0.25cm} c@{\hskip 0.25cm} c@{\hskip 0.25cm} c@{\hskip 0.25cm} c@{\hskip 0.25cm} c}
\hline \vspace{-0.2in} \\ 
& $X_1$ & $X_2$ & $X_3$ & $X_4$ & $X_5$ & $X_6$ & $X_7$ & $X_8$ & $X_9$ & $X_{10}$ & $X_{11}$ & $X_{12}$ & $X_{13}$ \\
\hline \vspace{-0.15in} \\
$X_1$ & \ 1.000 & \  \textcolor{blue}{0.954} & \ 0.364 & \ 0.342 & -0.129 & \ 0.313 & \ 0.496 & \ 0.424 & \ 0.592 & \ 0.545 & \ 0.084 & -0.019 & \ 0.134 \\
$X_2$ & \  \textcolor{blue}{0.954} & \ 1.000 & \ 0.297 & \ 0.284 & -0.118 & \ 0.291 & \ 0.503 & \ 0.419 & \ 0.648 & \ 0.569 & \ 0.076 & -0.036 & \ 0.144 \\
$X_3$ & \ 0.364 & \ 0.297 & \ 1.000 & \  \textcolor{blue}{0.882} & -0.148 & \ 0.153 & -0.029 & -0.054 & \ 0.125 & -0.081 & \ 0.162 & \ 0.220 & \ 0.126 \\
$X_4$ & \ 0.342 & \ 0.284 & \  \textcolor{blue}{0.882} & \ 1.000 & \ 0.220 & \ 0.381 & \ 0.174 & -0.059 & \ 0.137 & -0.014 & \ 0.097 & \ 0.169 & \ 0.015 \\
$X_5$ & -0.129 & -0.118 & -0.148 & \ 0.220 & \ 1.000 & \ 0.364 & \ 0.296 & \ 0.004 & -0.039 & \ 0.037 & -0.091 & -0.145 & -0.208 \\
$X_6$ & \ 0.313 & \ 0.291 & \ 0.153 & \ 0.381 & \ 0.364 & \ 1.000 & \  \textcolor{blue}{0.813} & \ 0.090 & \ 0.211 & \ 0.274 & -0.036 & \ 0.024 & -0.329 \\
$X_7$ & \ 0.496 & \ 0.503 & -0.029 & \ 0.174 & \ 0.296 & \  \textcolor{blue}{0.813} & \ 1.000 & \ 0.372 & \ 0.465 & \ 0.679 & -0.113 & -0.232 & -0.424 \\
$X_8$ & \ 0.424 & \ 0.419 & -0.054 & -0.059 & \ 0.004 & \ 0.090 & \ 0.372 & \ 1.000 & \ 0.482 & \ 0.557 & \ 0.061 & -0.357 & -0.202 \\
$X_9$ & \ 0.592 & \ 0.648 & \ 0.125 & \ 0.137 & -0.039 & \ 0.211 & \ 0.465 & \ 0.482 & \ 1.000 & \ 0.526 & \ 0.085 & -0.127 & -0.076 \\
$X_{10}$ & \ 0.545 & \ 0.569 & -0.081 & -0.014 & \ 0.037 & \ 0.274 & \ 0.679 & \ 0.557 & \ 0.526 & \ 1.000 & -0.319 & -0.368 & -0.291 \\
$X_{11}$ & \ 0.084 & \ 0.076 & \ 0.162 & \ 0.097 & -0.091 & -0.036 & -0.113 & \ 0.061 & \ 0.085 & -0.319 & \ 1.000 & \ 0.029 & \ 0.007 \\
$X_{12}$ & -0.019 & -0.036 & \ 0.220 & \ 0.169 & -0.145 & \ 0.024 & -0.232 & -0.357 & -0.127 & -0.368 & \ 0.029 & \ 1.000 & \ 0.184 \\
$X_{13}$ & \ 0.134 & \ 0.144 & \ 0.126 & \ 0.015 & -0.208 & -0.329 & -0.424 & -0.202 & -0.076 & -0.291 & \ 0.007 & \ 0.184 & \ 1.000 \\
				
\hline
\end{tabular}
\end{center}
\end{table}
\end{linespread}

Table \ref{corpitprop} presents the sample correlation matrix of the 13 physical variables for the pitprop data. The correlations between $X_1$ and $X_2$, between $X_3$ and $X_4$, and between $X_6$ and $X_7$ are strong. There are also a number of moderate correlations. The correlation matrix has the eigenvalues 4.219, 2.378, 1.878, 1.109,  0.910, 0.815, 0.576, 0.440, 0.353, 0.191, \textcolor{blue}{0.051},  \textcolor{blue}{0.041} and \textcolor{blue}{0.039}. Three of these are small compared to the others, suggesting that there are three collinear sets in the dataset. 

Table \ref{eigenpit} displays the eigenvectors corresponding to the three small eigenvalues along with the VIFs. The variables $X_1$, $X_2$, $X_3$, $X_4$, $X_6$, $X_7$ and $X_{10}$ have VIFs greater than 5, suggesting that each is involved in at least one collinearity. The eigenvectors $\boldsymbol{v}_{13}$ and $\boldsymbol{v}_{12}$, which are associated with the two smallest eigenvalues, both suggest a collinearity between $X_1$, $X_2$, $X_3$ and $X_4$. The condition index for the third smallest eigenvalue is 83, a little below the threshold of 100 that is generally used to indicate moderate collinearity. However, the threshold seems a little too high, as $X_7$ has a VIF of 12.4, so it is clearly involved in a collinearity, and that collinearity can only relate to the third smallest eigenvalue.  As well as $X_7$, the collinearity also involves $X_6$, and perhaps involves $X_{10}$, whose VIF is just above 5.

\begin{linespread}{1.5}
\begin{table}
\caption{\footnotesize Eigenvector loadings and VIFs for the pitprop data}\label{eigenpit} \vspace{-0.05 in}
\newcommand\Fontvi{\fontsize{9.0}{9.0}\selectfont}
\Fontvi
\begin{center}
\begin{tabular}{l@{\hskip 0.15cm} c@{\hskip 0.15cm} c@{\hskip 0.15cm} c@{\hskip 0.15cm} c@{\hskip 0.15cm} c@{\hskip 0.15cm} c@{\hskip 0.15cm} c@{\hskip 0.15cm} c@{\hskip 0.15cm} c@{\hskip 0.15cm} c@{\hskip 0.15cm} c@{\hskip 0.15cm} c@{\hskip 0.15cm} c }
\hline \vspace{-0.15in} \\ 
Eigenvector & $X_1$ & $X_2$ & $X_3$ & $X_4$ & $X_5$ & $X_6$ & $X_7$ & $X_8$ & $X_9$ & $X_{10}$ & $X_{11}$ & $X_{12}$ & $X_{13}$ \\
\hline \vspace{-0.15in} \\
$\boldsymbol{v}_{11}$ & \ 0.005 & \ 0.054 & -0.117 & \ 0.017 & \ -0.005 & \ \textcolor{blue}{0.537} & \ \textcolor{blue}{-0.764} & \ -0.026 & \ 0.051 & \ \textcolor{purple}{0.318} & \ 0.048 & -0.047 & -0.045 \\
$\boldsymbol{v}_{12}$ & \ \textcolor{blue}{0.392} & \textcolor{blue}{-0.411} & \ \textcolor{blue}{0.527} & \textcolor{blue}{-0.585} & \ 0.202 & \ 0.080 & -0.036 & -0.053 & \ 0.054 & \ 0.060 & \ 0.005 & \ 0.002 & \ 0.013 \\
$\boldsymbol{v}_{13}$ & \ \textcolor{blue}{0.572} & \textcolor{blue}{-0.582} & \textcolor{blue}{-0.408} & \ \textcolor{blue}{0.383} & -0.118 & -0.057 & -0.002 & -0.018 & \ 0.058 & -0.004 & \ 0.007 & -0.004 & \ 0.009 \\
\hline \vspace{-0.15 in} \\
VIF & 13.135 & 13.714 & 11.660 & 12.420 & \ 2.533 & \ 6.932 & 12.033 & \ 1.852 & \ 2.103 & \ 5.118 & \ 1.511 & \ 1.434 & \ 1.771 \\
\hline
\vspace{-.2in}
\end{tabular}
\end{center}
\end{table}
\end{linespread}

Table \ref{vdppitpror} presents the results obtained using the variance decomposition method for the five largest condition indices. The second column indicates that there are three weak collinearities (condition indices around 100). The last two condition indices are roughly equal. Belsley et al. [2] suggests that this indicates competing dependencies (collinearities) and that  VDPs should be combined over the rows of the condition indices that are similar and that it is not possible to identify the variables involved in each dependency. They write:
\begin{quote}
``Here involvement is determined by aggregating the variance-decomposition proportions over the competing condition indexes. Those variates whose {\em aggregate} proportions exceed the threshold $\pi^*$ are involved in at least one of the competing dependencies, and therefore may have degraded coefficient estimates. In this case, it is not possible exactly to determine in which of the competing near dependencies the variates are involved.''
\end{quote}
Belsley et al. [2] suggests a threshold $\pi^* =0.5$, under which the third largest condition index indicates a collinearity between $X_6$ and $X_7$. Combining the variance decomposition proportions over the last two rows, $X_1$, $X_2$, $X_3$, $X_4$ and $X_5$ are involved in collinearities. This differs a little from the VIFs, which suggest that $X_{10}$ is involved in a collinearity while $X_5$ is not.

\begin{linespread}{1.5} 
\begin{table}
\caption{Condition indices and variance-decomposition proportions for the pitprop data}\label{vdppitpror}\vspace{-0.05in}  
\newcommand\Fontvi{\fontsize{9.0}{9.0}\selectfont}
\Fontvi
\begin{center}
\begin{tabular}{ c@{\hskip 0.2cm} c@{\hskip 0.2cm} c@{\hskip 0.2cm} c@{\hskip 0.2cm} c@{\hskip 0.2cm} c@{\hskip 0.2cm} c@{\hskip 0.2cm} c@{\hskip 0.2cm} c@{\hskip 0.2cm} c@{\hskip 0.2cm} c@{\hskip 0.2cm} c@{\hskip 0.2cm} c@{\hskip 0.2cm} c@{\hskip 0.2cm} c c }
\hline \vspace{-0.2in} \\ 
Eigen- & Condition & \multicolumn{13}{c}{Variance-decomposition proportion} \\
\cline{3-15}
vector & Index & $X_1$ & $X_2$ & $X_3$ & $X_4$ & $X_5$ & $X_6$ & $X_7$ & $X_8$ & $X_9$ & $X_{10}$ & $X_{11}$ & $X_{12}$ & $X_{13}$  \\
\hline \vspace{-0.15in} \\
$\boldsymbol{v}_9$  & \ 12.6 & 0.024 & 0.022 & 0.001 & 0.000 & 0.088 & 0.069 & 0.004 & 0.183 & 0.195 & 0.040 & 0.042 & 0.013 & 0.236 \\
$\boldsymbol{v}_{10}$  & \ 22.1 & 0.038 & 0.028 & 0.002 & 0.004 & 0.000 & 0.008 & 0.015 & 0.016 & 0.008 & 0.518 & 0.394 & 0.138 & 0.318 \\
$\boldsymbol{v}_{11}$ & \ 83.4 & 0.000 & 0.004 & 0.023 & 0.000 & 0.000 & \textcolor{blue}{0.823} & \textcolor{blue}{0.959} & 0.007 & 0.024 & 0.391 & 0.031 & 0.031 & 0.022 \\
$\boldsymbol{v}_{12}$  & 101.7 & \textcolor{blue}{0.282} & \textcolor{blue}{0.297} & \textcolor{blue}{0.574} & \textcolor{blue}{0.665} & \textcolor{blue}{0.390} & 0.022 & 0.003 & 0.036 & 0.033 & 0.017 & 0.000 & 0.000 & 0.002 \\
$\boldsymbol{v}_{13}$  & 108.9 & \textcolor{blue}{0.643} & \textcolor{blue}{0.638} & \textcolor{blue}{0.368} & \textcolor{blue}{0.305} & \textcolor{blue}{0.143} & 0.012 & 0.000 & 0.005 & 0.041 & 0.000 & 0.001 & 0.000 & 0.001 \\
\hline 
\end{tabular}
\end{center}
\end{table}
\end{linespread}

Auxiliary regressions are often fitted as part of the variance-decomposition analysis when there are   two or more collinearities. From each row with a large condition index, a variable is selected that has a large VDP in that row and comparatively small VDPs in the other rows with large condition indices. Each selected variable is used in turn as the dependent variable in a regression analysis to identify the variables with which it is collinear  [1]. From the middle row of Table \ref{vdppitpror},  $X_7$, is selected as it has the largest VDP in that row and very small VDPs in the lowest two rows of the table, which are the other rows with large condition indices. For similar reasons, $X_4$ is selected from the row with a condition index of 101.7 and $X_1$ from the row with a condition index of 108.9. 

When a stepwise regression is performed with $X_7$ as the dependent variable, variables are entered into the regression equation in the order $X_6$ (0.659), $X_{10}$ (0.885), $X_3$ (0.893), $X_9$ (0.9014). The figures in brackets are the $R^2$ values after each variable has been added to the regression; as they only increase slowly after $X_6$ and $X_{10}$ have been entered into the regression, the analysis indicates a collinearity between $X_7$,  $X_6$ and $X_{10}$. (The effect of adding $X_9$ is actually significant at the 0.001 significance level, but the improvement in $R^2$ is small and, moreover, the VIF of $X_9$ is small and does not suggest that it is involved in a collinearity.) When $X_4$ is the dependent variable, explanatory variables are entered in the order $X_3$ (0.777), $X_{5}$ (0.903), $X_6$ (0.916), suggesting a collinearity between $X_4$, $X_3$ and $X_{5}$, even though $X_{5}$ has a VIF that is well below the threshold value of 5. With $X_1$ as the dependent variable, variables are entered in the order $X_2$ (0.910), $X_{3}$ (0.916), $X_8$ (0.918), suggesting a collinearity between $X_1$ and $X_2$. 

Turning to the cos-max method, Table \ref{maxpit} gives the transformation matrix for the pitprop data, $\bA = (\ba_1, \ldots. \ba_{13})$, together with the VIFs, which equal $\ba_1^T \ba_1, \ldots,\ba_{13}^T \ba_{13}$. Treating VIFs above 5 as indicative of collinearity, we examine the components of $\ba_1$, $\ba_2$, $\ba_3$, $\ba_4$, $\ba_6$, $\ba_7$ and $\ba_{10}$. Looking at $\ba_1^T$, its first two components are large (well above a threshold of 0.75) while its other components are small, indicating a collinearity between $X_1$ and $X_2$. The components of $\ba_2^T$ identify the same collinearity. In $\ba_3^T$, the components for $X_3$ and $X_4$ are well above 0.75, that for $X_5$ is a little below that threshold, and other components are well below the threshold. This indicates a collinearity between $X_3$ and $X_4$ that perhaps involves $X_5$. The components of $\ba_4^T$ give the same inference. Looking at $\ba_6^T$, $\ba_7^T$ and $\ba_{10}^T$, a collinearity between $X_6$ and $X_7$ is suggested by $\ba_6^T$, a collinearity between $X_6$, $X_7$ and $X_{10}$ is suggested by $\ba_7^T$, and  a collinearity between $X_6$, $X_7$ and $X_{10}$ is suggested by $\ba_{10}^T$. The VIF from  $\ba_7^T$ is much bigger than the VIF from  $\ba_6^T$ or  $\ba_{10}^T$, suggesting that the collinearity is more clearly defined by  $\ba_7^T$  than by  $\ba_6^T$ or  $\ba_{10}^T$. (Interestingly, the VIF of  $\ba_7^T$ almost exactly equals the sum of the VIFs of  $\ba_6^T$ and  $\ba_{10}^T$.) This implies that there is a collinearity between $X_6$, $X_7$ and $X_{10}$. This example illustrates that the cos-max method considers far more information to identify collinearities than an eigenvector analysis, or a variance-decomposition analysis that does not include auxiliary regressions.

\begin{linespread}{1.5} 
\begin{table}
\caption{\footnotesize Cos-max transformation matrix and VIFs for the pitprop data}\label{maxpit} \vspace{-0.05in} 
\newcommand\Fontvi{\fontsize{9.0}{9.0}\selectfont}
\Fontvi
\begin{center}
\begin{tabular}{ l@{\hskip 0.15cm} c@{\hskip 0.15cm} c@{\hskip 0.15cm} c@{\hskip 0.15cm} c@{\hskip 0.15cm} c@{\hskip 0.15cm} c@{\hskip 0.15cm} c@{\hskip 0.15cm} c@{\hskip 0.15cm} c@{\hskip 0.15cm} c@{\hskip 0.15cm} c@{\hskip 0.15cm} c@{\hskip 0.15cm} c@{\hskip 0.15cm} c}
\hline \vspace{-0.15in} \\ 
& $X_1$ & $X_2$ & $X_3$ & $X_4$ & $X_5$ & $X_6$ & $X_7$ & $X_8$ & $X_9$ & $X_{10}$ & $X_{11}$ & $X_{12}$ & $X_{13}$ & VIF  \\
\hline \vspace{-0.15in} \\
$\boldsymbol{a}_1^{\top}$ & \ \textcolor{blue}{3.036} & \textcolor{blue}{-1.912} & -0.252 & -0.100 & \ 0.119 & -0.064 & -0.220 & -0.202 & \ 0.006 & -0.204 & -0.079 & -0.068 & -0.175 & 13.135 \\ 
$\boldsymbol{a}_2^{\top}$ & \textcolor{blue}{-1.912} & \ \textcolor{blue}{3.114} & \ 0.004 & -0.051 & \ 0.048 & \ 0.032 & -0.220 & \ 0.004 & -0.382 & -0.292 & -0.103 & -0.059 & -0.255 & 13.714 \\
$\boldsymbol{a}_3^{\top}$ & -0.252 & \ 0.004 & \ \textcolor{blue}{2.717} & \textcolor{blue}{-1.946} & \ \textcolor{purple}{0.592} & \ 0.017 & \ 0.259 & -0.027 & -0.015 & \ 0.077 & -0.053 & -0.028 & \ 0.022 & 11.660 \\
$\boldsymbol{a}_4^{\top}$ & -0.100 & -0.051 & \textcolor{blue}{-1.946} & \ \textcolor{blue}{2.838} & \textcolor{purple}{-0.664} & -0.333 & \ 0.020 & \ 0.116 & -0.032 & -0.015 & \ 0.007 & -0.041 & \ 0.011 & 12.420 \\
$\boldsymbol{a}_5^{\top}$ & \ 0.119 & \ 0.048 & \ \textcolor{purple}{0.592} & \textcolor{purple}{-0.664} & \ 1.296 & -0.108 & -0.128 & -0.040 & \ 0.018 & \ 0.071 & \ 0.036 & \ 0.086 & \ 0.017 & \ 2.533 \\
$\boldsymbol{a}_6^{\top}$ & -0.064 & \ 0.032 & \ 0.017 & -0.333 & -0.108 & \ \textcolor{blue}{2.102} & \textcolor{blue}{-1.479} & \ 0.043 & \ 0.098 & \ 0.394 & \ 0.037 & -0.154 & \ 0.069 & \ 6.932 \\
$\boldsymbol{a}_7^{\top}$ & -0.220 & -0.220 & \ 0.259 & \ 0.020 & -0.128 & \textcolor{blue}{-1.479} & \ \textcolor{blue}{2.979} & \ 0.004 & -0.211 & \textcolor{blue}{-0.788} & -0.023 & \ 0.170 & \ 0.305 & 12.033 \\
$\boldsymbol{a}_8^{\top}$ & -0.202 & \ 0.004 & -0.027 & \ 0.116 & -0.040 & \ 0.043 & \ 0.004 & \ 1.288 & -0.171 & -0.254 & -0.095 & \ 0.149 & \ 0.094 & \ 1.852 \\
$\boldsymbol{a}_9^{\top}$ & \ 0.006 & -0.382 & -0.015 & -0.032 & \ 0.018 & \ 0.098 & -0.211 & -0.171 & \ 1.360 & -0.124 & -0.077 & -0.021 & \ 0.017 & \ 2.103 \\
$\boldsymbol{a}_{10}^{\top}$ & -0.204 & -0.292 & \ 0.077 & -0.015 & \ 0.071 & \ 0.394 & \textcolor{blue}{-0.788} & -0.254 & -0.124 & \ \textcolor{blue}{1.969} & \ 0.411 & \ 0.191 & \ 0.207 & \ 5.118 \\
$\boldsymbol{a}_{11}^{\top}$ & -0.079 & -0.103 & -0.053 & \ 0.007 & \ 0.036 & \ 0.037 & -0.023 & -0.095 & -0.077 & \ 0.411 & \ 1.137 & \ 0.056 & \ 0.097 & \ 1.511 \\
$\boldsymbol{a}_{12}^{\top}$ & -0.068 & -0.059 & -0.028 & -0.041 & \ 0.086 & -0.154 & \ 0.170 & \ 0.149 & -0.021 & 0.191 & \ 0.056 & \ 1.141 & \ 0.000 & \ 1.434 \\
$\boldsymbol{a}_{13}^{\top}$ & -0.175 & -0.255 & \ 0.022 & \ 0.011 & \ 0.017 & \ 0.069 & \ 0.305 & \ 0.094 & \ 0.017 & \ 0.207 & \ 0.097 & \ 0.000 & \ 1.231 & \ 1.771 \\
\hline 
\end{tabular}
\end{center}
\end{table}
\end{linespread}

In this example, the superiority of the cos-max method over eigenvector analysis is that the eigenvector analysis could not suggest two separate collinearities, one between $X_1$ and $X_2$ and another between $X_3$ and $X_4$, while the cos-max method identifies these two non-overlapping collinearities. Similarly, without using auxiliary regressions, the variance-decomposition method could only indicate that the variables $X_1$, $X_2$, $X_3$, $X_4$ and $X_5$ are involved in at least one of the collinearities. The auxiliary regressions and the cos-max method identified similar collinearities -- the collinearities would be identical if the cut-off for the cos-max method was relaxed from 0.75 to 0.59, when $X_5$ would be added to a collinearity of the cos-max method. Hence the two substantially different methods largely support each other's inferences.

When there is more than one collinearity, there are many ways these collinearities may be defined. For example, suppose two collinearities are specified as $\sum b_i X_i$ and $\sum c_i X_i$ (so  $\sum b_i \bx_i \approx 0$ and $\sum c_i \bx_i \approx 0$, where the $b_i$ and $c_i$ are scalar coefficients). Then $\sum ( \gamma  b_i + \eta c_i)X_i \approx 0$ for any scalars $\gamma$ and $\eta$. Hence the collinearities could be replaced with collinearities of the form  $\sum ( \gamma _1 b_i + \eta _1 c_i)X_i$ and $\sum ( \gamma _2 b_i + \eta _2 c_i)X_i$, where the only restriction on the constants $\gamma_1$,  $\eta_1$, $\gamma_2$ and $\eta_2$ is that the new collinearities must not be linearly dependent. However, simplicity is desirable so it is better to specify collinearities whose variables do not overlap, when possible. Thus, with the pitprop data it is more informative to say there is a collinearity between $X_1$ and $X_2$ and a second collinearity between $X_3$ and $X_4$, rather than say there are two collinearities between $X_1$, $X_2$, $X_3$ and $X_4$. More generally, it may be necessary to have some overlap in the variables involved in different collinearites, but keeping the overlap to a minimum should typically give a clearer understanding of the data. This is relevant to the next example.

\subsubsection*{\\ 3.3 Example 3: Shopping Pattern Data (multiple collinearities)} \label{S:Ex3}

This example concerns data used by Mahajan et al. [8] to illustrate the application of ridge regression to collinear data. The data related to the convenience, quality,
 and service dimensions of food retail outlets and were collected through telephone interviews from people who  did the major food shopping for their household. There were 10 regressors in their study: distance from the store ($X_1$), size of the parking lot ($X_2$), outside appearance of the store ($X_3$), sales area within the store ($X_4$), number of aisles ($X_5$), number of checkout counters ($X_6$), inside appearance of the store ($X_7$), quality of meat department ($X_8$), quality of produce department ($X_9$), and quality of grocery department ($X_{10}$). Table \ref{corbus} gives the correlation matrix of the regressors. There are a number of strong correlations, such as between $X_2$ and $X_4$, between $X_5$ and $X_6$ and between $X_9$ and $X_{10}$, as well as a number of moderate correlations among the regressors. Large pairwise correlation is a sufficient condition for collinearity, so we expect collinearities between some of the regressors. 
 
\begin{linespread}{1.5}
\begin{table}
\caption{Correlation matrix for the regressors in the shopping pattern data} \label{corbus} \vspace{-0.05in}
\newcommand\Fontvi{\fontsize{10.0}{10.0}\selectfont}
\Fontvi
\begin{center}
\begin{tabular}{c c@{\hskip 0.35cm} c@{\hskip 0.35cm} c@{\hskip 0.35cm} c@{\hskip 0.35cm} c@{\hskip 0.35cm} c@{\hskip 0.35cm} c@{\hskip 0.35cm} c@{\hskip 0.35cm} c@{\hskip 0.35cm} c}
\hline \vspace{-0.2in} \\ 
& $X_1$ & $X_2$ & $X_3$ & $X_4$ & $X_5$ & $X_6$ & $X_7$ & $X_8$ & $X_9$ & $X_{10}$  \\
\hline \vspace{-0.15in} \\
$X_1$ & 1.000 & 0.547 & 0.274 & 0.637 & 0.481 & 0.517 & 0.369 & 0.242 & 0.566 & 0.666 \\
$X_2$ & 0.547 & 1.000 & 0.650 & \textcolor{blue}{0.837} & 0.809 & 0.792 & 0.562 & 0.277 & 0.808 & 0.768 \\
$X_3$ & 0.274 & 0.650 & 1.000 & 0.744 & 0.782 & 0.795 & 0.781 & 0.496 & 0.566 & 0.526 \\
$X_4$ & 0.637 & \textcolor{blue}{0.837} & 0.744 & 1.000 & 0.851 & 0.772 & 0.609 & 0.609 & 0.815 & 0.775 \\
$X_5$ & 0.481 & 0.809 & 0.782 & 0.851 & 1.000 & \textcolor{blue}{0.906} & 0.821 & 0.573 & 0.798 & 0.748 \\
$X_6$ & 0.517 & 0.792 & 0.795 & 0.772 & \textcolor{blue}{0.906} & 1.000 & 0.781 & 0.418 & 0.736 & 0.702 \\
$X_7$ & 0.369 & 0.562 & 0.781 & 0.609 & 0.821 & 0.781 & 1.000 & 0.473 & 0.577 & 0.599 \\
$X_8$ & 0.242 & 0.277 & 0.496 & 0.609 & 0.573 & 0.418 & 0.473 & 1.000 & 0.484 & 0.424 \\
$X_9$ & 0.566 & 0.808 & 0.566 & 0.815 & 0.798 & 0.736 & 0.577 & 0.484 & 1.000 & \textcolor{blue}{0.894} \\
$X_{10}$ & 0.666 & 0.768 & 0.526 & 0.775 & 0.748 & 0.702 & 0.599 & 0.424 & \textcolor{blue}{0.894} & 1.000 \\

\hline
\vspace{-.4in}
\end{tabular}
\end{center}
\end{table}
\end{linespread} 

The eigenvalues of the correlation matrix are 6.897, 1.059, 0.739, 0.462, 0.349, 0.188, 0.129, \textcolor{blue}{0.081}, \textcolor{blue}{0.064} and \textcolor{blue}{0.032}. The corresponding condition indices are 1.0, 6.5, 9.3, 14.9, 19.8, 36.7, 53.6, \textcolor{blue}{84.7}, \textcolor{blue}{108.0} and \textcolor{blue}{216.6}, respectively. The last three eigenvalues are small compared to the others and hence give three large condition indices, suggesting three collinear sets in the data. 

Table \ref{eigenshop} presents the eigenvectors and condition indices corresponding to the three smallest eigenvalues, together with the VIFs of the regressors. The VIFs indicate that all the regressors except $X_1$ and $X_8$ seem to be involved in a collinearity. From $\boldsymbol{v}_8$, the eigenvector corresponding to the third smallest eigenvalue, there is a collinearity between $X_9$ and $X_{10}$ and,  from $\boldsymbol{v}_9$, another collinearity is between $X_2$, $X_4$, and $X_5$. From  $\boldsymbol{v}_{10}$, the third collinearity clearly involves $X_4$ and $X_5$. Perhaps $X_3$ and/or $X_6$ are also involved in that collinearity, as that is the only collinearity in which they could be involved and they have quite large VIFs. The only variable with a VIF above a threshold of 5 that is not included in these collinearities is $X_7$, and its VIF is only slightly above 5.

\begin{linespread}{1.5}
\begin{table}
\caption{\footnotesize Eigenvectors analysis, condition indices and VIFs for the shopping pattern data}\label{eigenshop} \vspace{-0.05 in}
\newcommand\Fontvi{\fontsize{10.0}{10.0}\selectfont}
\Fontvi
\begin{center}
\begin{tabular}{l@{\hskip 0.15cm} c@{\hskip 0.15cm} c@{\hskip 0.15cm} c@{\hskip 0.15cm} c@{\hskip 0.15cm} c@{\hskip 0.15cm} c@{\hskip 0.15cm} c@{\hskip 0.15cm} c@{\hskip 0.15cm} c@{\hskip 0.15cm} c@{\hskip 0.15cm} c@{\hskip 0.15cm}  c }
\hline \vspace{-0.15in} \\ 
Eigen- & Eigen- & Cond. & \multicolumn{10}{c}{Eigenvector loadings} \\
\cline{4-13}  \vspace{0.02in}
vector & value & index & $X_1$ & $X_2$ & $X_3$ & $X_4$ & $X_5$ & $X_6$ & $X_7$ & $X_8$ & $X_9$ & $X_{10}$  \\
\hline \vspace{-0.15in} \\
$\boldsymbol{v}_8$ &   0.081 & \  84.7 & \ 0.110 & -0.247 & \ 0.035 & \ 0.151 & \ 0.041 & -0.186 & \ 0.146 & -0.178 &  \ \ \textcolor{blue}{0.683} & \ \textcolor{blue}{-0.589} \\
$\boldsymbol{v}_9$ & 0.064 & 108.0 & -0.126 & \textcolor{blue}{-0.525} & -0.073 & \ \textcolor{blue}{0.478} & \ \textcolor{blue}{0.526} & -0.087 & -0.151 & -0.294 & \ -0.185 & \ \ 0.216 \\
$\boldsymbol{v}_{10}$ & 0.032 & 216.6 &\ 0.213 & \ 0.070 & \ \textcolor{purple}{0.337} & \textcolor{blue}{-0.554} & \ \textcolor{blue}{0.566} & \textcolor{purple}{-0.356} & -0.280 & \ 0.039 & \ \ 0.037 & \ \ 0.001 \\
\hline \vspace{-0.15 in} \\
& \multicolumn{2}{c}{VIFs:}  & \ 3.330 & \ 7.526 & \ 6.783 & 14.641 & 15.831 & \ \ 8.832 &  \ 5.875 & \ \ 2.947 & \ \  7.488 & \ \ 6.643 \\
\hline 
\end{tabular}
\end{center}
\end{table}
\end{linespread}

The variance-decomposition proportions from the shopping pattern data are given for the three largest condition indices in Table \ref{locsp}.  The top row in the table clearly indicates that $X_9$ and $X_{10}$ are collinear, which is one of the finding from the eigenvector analysis. The last row  indicates that $X_3$, $X_4$ and $X_5$ are involved in another collinearity, and that perhaps $X_6$ is involved in this set. This collinear set is the same as the collinear set identified by the last eigenvector ($\boldsymbol{v}_{10}$). The middle row has a condition index of 108.0, which strongly indicates a collinearity. However, in that row  only $X_2$ has a VDP that exceeds 0.5 and the only other variable with a VDP close to 0.5 is $X_8$. As a collinearity must involve at least two variables, the VDP analysis indicates a collinearity between $X_2$ and $X_8$. This is surprising because $X_8$ has a small VIF of only 2.947, well below a threshold of 5. Instead of a collinearity between $X_2$ and $X_8$, the eigenvector analysis suggested a collinearity between $X_2$, $X_4$ and $X_5$.

Auxiliary regressions were performed to explore the collinearities further. The variable $X_9$ has the largest VDP in the first row of Table \ref{locsp} and very small VDPs in the other two rows, so clearly it should be the dependent variable in one of the auxiliary regressions. For similar reasons, $X_2$ should be the dependent variable in a second auxiliary regression as it has the largest VDP in the second row of the table and small VDPs in the other two rows. The bottom row of the table suggests that either $X_3$ or $X_4$ should be taken as the dependent variable in the third auxiliary regression; other regressors are all poorer than $X_3$ or $X_4$ in terms of having a high VDP in the third row and low VDPs in the other two rows. 

\begin{linespread}{1.5} 
\begin{table}
\caption{Condition indices and variance-decomposition proportions for the shopping pattern data} \label{locsp}\vspace{-0.05in}
\newcommand\Fontvi{\fontsize{10.0}{10.0}\selectfont}
\Fontvi
\begin{center}
\begin{tabular}{ c@{\hskip 0.35cm} c@{\hskip 0.35cm} c@{\hskip 0.35cm} c@{\hskip 0.35cm} c@{\hskip 0.35cm} c@{\hskip 0.35cm} c@{\hskip 0.35cm} c@{\hskip 0.35cm} c@{\hskip 0.35cm} c@{\hskip 0.35cm} c@{\hskip 0.35cm}  cc@{\hskip 0.35cm} c }
\hline \vspace{-0.2in} \\ 
Condition & \multicolumn{10}{c}{Variance-decomposition proportion} \\
\cline{2-11}
index & $X_1$ & $X_2$ & $X_3$ & $X_4$ & $X_5$ & $X_6$ & $X_7$ & $X_8$ & $X_9$ & $X_{10}$  \\
\hline \vspace{-0.15in} \\
 \ 84.7 &  0.044 & 0.100 & 0.002 & 0.019 & 0.001 & 0.048 & 0.044 & 0.131 & \textcolor{blue}{0.766} & \textcolor{blue}{0.641} \\
108.0 &  0.074 & \textcolor{blue}{0.573} & 0.012 & 0.245 & 0.274 & 0.013 & 0.061 & \textcolor{purple}{0.460} & 0.071 & 0.110 \\
216.6  & 0.428 & 0.021 & \textcolor{blue}{0.527} & \textcolor{blue}{0.659} & \textcolor{blue}{0.636} & \textcolor{purple}{0.451} & 0.419 & 0.016 & 0.006 & 0.000 \\
\hline
\end{tabular}
\end{center}
\end{table}
\end{linespread}

With $X_9$ as the dependent variable, stepwise regression entered variables in the order $X_{10}$ (0.7966), $X_5$ (0.8335) and $X_7$ (0.8466). As before, the figures in brackets are the $R^2$ values after each variable has been added to the regression. They indicate a collinearity between $X_9$ and $X_{10}$. With $X_2$ as the dependent variable, other variables entered the regression in the order $X_4$ (0.6979), $X_8$ (0.7829), $X_5$ (0.8329) and $X_9$ (0.8511), suggesting a collinearity between $X_2$, $X_4$ and $X_8$. (Adding $X_5$ increases the $R^2$ value by only a modest amount.) With $X_4$ as the dependent variable, regressors entered in the order $X_5$ (0.7210), $X_1$ (0.7870) and $X_2$ (0.819), suggesting a collinearity between $X_4$, $X_5$ and $X_1$. With $X_3$ as the dependent variable, regressors entered in the order $X_6$ (0.6284), $X_7$ (0.6916) and $X_4$ (0.7301), suggesting a weak collinearity between $X_3$ and $X_6$ that perhaps involves $X_7$. Regardless of whether $X_3$ or $X_4$ is chosen as the dependent variable in the third auxiliary regression, at least one regressor with a large VIF is omitted from all collinearities: if $X_4$ is taken as the dependent variable in the third collinearity, then $X_3$, $X_6$ and $X_7$ are omitted from all the identified collinearities, while if $X_3$ is taken as the third dependent variable, then $X_5$ is omitted from all collinearities, even though $X_5$ has the largest VIF of all the regressors. This example illustrates that auxiliary equations may give results that are not fully coherent with VIFs.

Tables \ref{maxshop} presents the transformation matrices of the cos-max method for the shopping pattern data. The components of $\boldsymbol{a}_9$ and $\boldsymbol{a}_{10}$ clearly identify the collinearity between $X_9$ and $X_{10}$ that was found using the eigenvector analysis and the variance-decomposition method. Further information about the structure between $X_2$, $X_3$, $X_4$, $X_5$, $X_6$ and $X_7$ is provided by the components of $\boldsymbol{a}_2$, $\boldsymbol{a}_3$, $\boldsymbol{a}_4$, $\boldsymbol{a}_5$, $\boldsymbol{a}_6$ and $\boldsymbol{a}_7$. From $\boldsymbol{a}_2$, there is a suggestion that $X_2$ and $X_4$ are related. The components of $\boldsymbol{a}_3$ indicate a relationship between $X_3$ and $X_4$. The large values in $\boldsymbol{a}_4$ indicate that $X_4$ is related to $X_2$, $X_3$, and $X_5$. Similarly, $\boldsymbol{a}_5$ indicates that $X_5$ is related to $X_4$, $X_6$ and $X_7$.  A relationship between $X_6$ and $X_5$ is indicated by $\boldsymbol{a}_6$, and a relationship between $X_7$ and $X_5$ is indicated by $\boldsymbol{a}_7$. This analysis has examined eight $10 \times 1$ vectors ($\boldsymbol{a}_2$, $\boldsymbol{a}_3$, $\boldsymbol{a}_4$,  $\boldsymbol{a}_5$, $\boldsymbol{a}_6$, $\boldsymbol{a}_7$, $\boldsymbol{a}_9$ and 
$\boldsymbol{a}_{10}$), and hence utilised much more information than the eigenvector analysis or the variance-decomposition method, which each only examined three $10 \times 1$ vectors. 

\begin{linespread}{1.5} 
\begin{table}
\caption{Cos-max transformation matrix and VIFs for the shopping pattern data} \label{maxshop} \vspace{-0.2in}
\newcommand\Fontvi{\fontsize{10.0}{10.0}\selectfont}
\Fontvi
\begin{center}
\begin{tabular}{ l@{\hskip 0.35cm} c@{\hskip 0.35cm} c@{\hskip 0.35cm} c@{\hskip 0.35cm} c@{\hskip 0.35cm} c@{\hskip 0.35cm} c@{\hskip 0.35cm} c@{\hskip 0.35cm} c@{\hskip 0.35cm} c@{\hskip 0.35cm} c@{\hskip 0.35cm} c }
\hline \vspace{-0.2in} \\ 
& $X_1$ & $X_2$ & $X_3$ & $X_4$ & $X_5$ & $X_6$ & $X_7$ & $X_8$ & $X_9$ & $X_{10}$ & VIF  \\
\hline \vspace{-0.15in} \\
$\boldsymbol{a}_1^{\top}$ & \ 1.501 & \ 0.030 & \ 0.410 & -0.712 & \ 0.282 & -0.348 & -0.177 & \ 0.095 & \ 0.080 & -0.395 & \ 3.330 \\
$\boldsymbol{a}_2^{\top}$ & \ 0.030 & \ \textcolor{blue}{2.428} & -0.125 & \textcolor{blue}{-0.843} & -0.506 & -0.336 & \ 0.170 & \ 0.533 & -0.392 & -0.267 & \ 7.526 \\
$\boldsymbol{a}_3^{\top}$ & \ 0.410 & -0.125 & \ \textcolor{blue}{2.179} & \textcolor{blue}{-0.921} & \ 0.241 & -0.652 & -0.695 & -0.071 & \ 0.165 & \ 0.078 & \ 6.783 \\
$\boldsymbol{a}_4^{\top}$ & -0.712 & \textcolor{blue}{-0.843} & \textcolor{blue}{-0.921} & \ \textcolor{blue}{3.330} & \textcolor{blue}{-0.794} & \ 0.316 & \ 0.502 & -0.622 & -0.330 & -0.104 & 14.641 \\
$\boldsymbol{a}_5^{\top}$ & \ 0.282 & -0.506 & \ 0.241 & \textcolor{blue}{-0.794} & \ \textcolor{blue}{3.548} & \textcolor{blue}{-1.085} & \textcolor{blue}{-0.914} & -0.353 & -0.286 & -0.030 & 15.831 \\
$\boldsymbol{a}_6^{\top}$ & -0.348 & -0.336 & -0.652 & \ 0.316 & \textcolor{blue}{-1.085} & \ \textcolor{blue}{2.611} & -0.160 & \ 0.102 & -0.215 & \ 0.006 & \ 8.832 \\
$\boldsymbol{a}_7^{\top}$ & -0.177 & \ 0.170 & -0.695 & \ 0.502 & \textcolor{blue}{-0.914} & -0.160 & \ \textcolor{blue}{2.030} & -0.093 & \ 0.066 & -0.289  & \ 5.875 \\
$\boldsymbol{a}_8^{\top}$ & \ 0.095 & \ 0.533 & -0.071 & -0.622 & -0.353 & \ 0.102 & -0.093 & \ 1.443 & -0.182 & -0.039 & \ 2.947 \\
$\boldsymbol{a}_9^{\top}$ & \ 0.080 & -0.392 & \ 0.165 & -0.330 & -0.286 & -0.215 & \ 0.066 & -0.182 & \ \textcolor{blue}{2.459} & \textcolor{blue}{-0.989} & \ 7.488 \\
$\boldsymbol{a}_{10}^{\top}$ & -0.395 & -0.267 & \ 0.078 & -0.104 & -0.030 & \ 0.006 & -0.289 & -0.039 & \textcolor{blue}{-0.989} & \ \textcolor{blue}{2.310} & \ 6.643 \\
\hline \vspace{-.4in}
\end{tabular}
\end{center}
\end{table}
\end{linespread} 

\begin{figure} 
\centering
\includegraphics[scale=0.9]{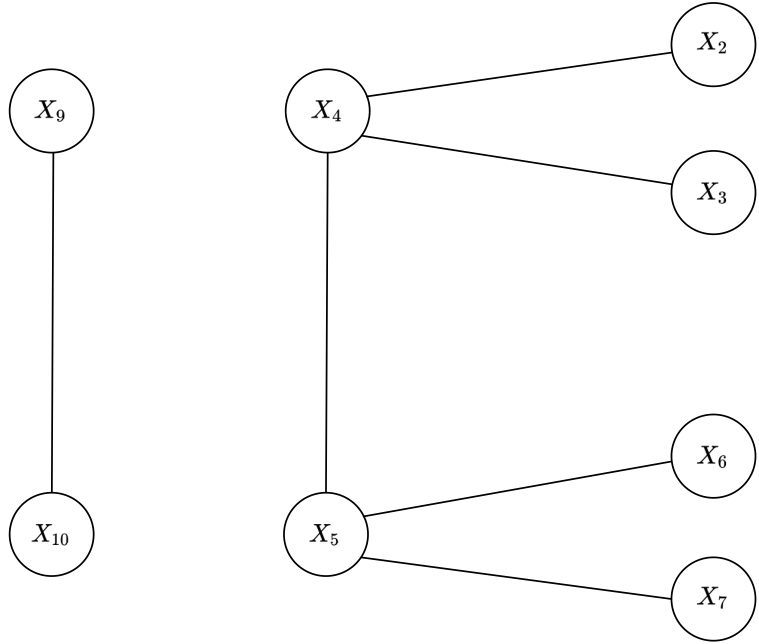}
\caption{Links between variables identified by the cos-max method}\label{multstructure} 
\end{figure}

The links between regressors that the cos-max method has identified are summarised in Figure \ref{multstructure}. It shows the simple relationship between $X_9$ and $X_{10}$ and the more complex relationship between $X_2$, $X_3$, $X_4$, $X_5$, $X_6$ and $X_7$. The top right-hand part of the figure suggests that $X_2$ and $X_3$ are related through their relationship with $X_4$ and this is indeed the case: the raw correlation between $X_2$ and $X_3$ is 0.650 (c.f. Table \ref{corbus})  while their partial correlation, conditional on $X_4$, is only 0.073. Similarly, the correlation between $X_6$ and $X_7$ is 0.781 and this reduces substantially when conditioned on $X_5$, to a partial correlation of 0.153, largely in line with the lower right-hand part of Figure \ref{multstructure}.

There are three small eigenvalues. One of these relates to the collinearity between $X_9$ and $X_{10}$. The other two indicate that variation in the six variables that are linked in Figure \ref{multstructure} ($X_2$, $X_3$, $X_4$, $X_5$, $X_6$ and $X_7$) is largely contained in a hyperplane whose dimension equals 6-2, i.e. in a four-dimensional hyperplane. This, and the links shown in Figure  \ref{multstructure}, should perhaps be all that is inferred from the cos-max method about the relationships between $X_2$, $X_3$, $X_4$, $X_5$, $X_6$ and $X_7$. If two collinearities from these variables need to be specified, then one would include the variables $X_4$, $X_2$ and $X_3$, and the other would include the variables $X_5$, $X_6$ and $X_7$. Also, as $X_4$ and $X_5$ are linked, at least one of these collinearities should contain both $X_4$ and $X_5$. Thus the first of them should include $X_5$ and/or the second one should include $X_4$. Partial correlations can inform the choice. Conditional on $X_4$, the partial correlation of $X_5$ with $X_2$ is 0.337 and the partial correlation of $X_5$ with $X_3$ is 0.425. These are not small, so $X_5$ should be included in the collinearity that contains $X_2$, $X_3$ and $X_4$. In contrast, conditional on $X_5$, the partial correlations of $X_4$ with $X_6$ and with $X_7$ are 0.005 and -0.298, respectively. These are quite small, suggesting that $X_4$ need not be included in the collinearity containing $X_5$, $X_6$ and $X_7$. To summarise, Figure \ref{multstructure} and the partial correlations together suggest that \{$X_2, \, X_3, \, X_4, \, X_5$\}, \{$X_5, \, X_6, \, X_7$\} and \{$X_9, \, X_{10}$\} form a reasonable set of three collinearities. They form a parsimonious choice, as together they include all the regressors with a VIF above 5 and the only overlap between them is that two collinearities contain $X_5$.

The question might still be asked of whether the structure shown in Figure \ref{multstructure} is a meaningful reflection of real structure in the data. Factor analysis was used to further explore the relationships between $X_2$, $X_3$, $X_4$, $X_5$, $X_6$ and $X_7$. There was clear evidence that a model with 2 factors did not adequately fit the correlations between these six variables ($\chi ^2 = 25.49$ on 4 degrees of freedom; $p <0.00005$). The model with three factors has 0 degrees of freedom for testing model adequacy and accounts for 89.4\% of the variation in the data. Table \ref{loadings}  shows the factor loadings of variables for the three-factor model fitted with varimax rotation. Moderately high loadings (between 0.5 and 0.75) are typed in purple and high loadings (above 0.75) are typed in blue.

\begin{linespread}{1.5} 
\begin{table}
\caption{Factor loadings of variables from a three-factor model fitted with varimax rotation} \label{loadings} \vspace{-0.05in}
\newcommand\Fontvi{\fontsize{10.0}{10.0}\selectfont}
\Fontvi
\begin{center}
\begin{tabular}{ l c c c c c c c  }
\hline \vspace{-0.2in} \\ 
& $X_2$ & $X_3$ & $X_4$ & $X_5$ & $X_6$ & $X_7$ \\
\hline \vspace{-0.15in} \\
Factor 1 & \textcolor{blue}{0.811} & 0.400 & \textcolor{blue}{0.817} & \textcolor{purple}{0.678} & \textcolor{purple}{0.599} & 0.289 \\
Factor 2 & 0.302 & 0.457 & 0.302 & \textcolor{purple}{0.673} & \textcolor{purple}{0.609} & \textcolor{blue}{0.780} \\
Factor 3 & 0.237 & \textcolor{blue}{0.791} & 0.352 & 0.257 & 0.350 & 0.390 \\
		
\hline \vspace{-.4in}
\end{tabular}
\end{center}
\end{table}
\end{linespread} 

Comparison of Table \ref{loadings} with the right-hand part of Figure \ref{multstructure} yields the following points.
\begin{enumerate}
\item Variables that are not linked in Figure \ref{multstructure} do not have the same pattern of loadings on factors.
\item Some variables that are linked in Figure \ref{multstructure} have the same pattern of loadings on factors. Specifically, $X_2$ and $X_4$ both have high loadings on Factor 1 and low loadings on Factors 2 and 3; $X_5$ and $X_6$ have moderately high loadings on Factors 1 and 2 and low loadings on Factor 3.
\item The pattern of loadings for $X_3$ (high on Factor 3 and low on Factors 1 and 2) has more in common with the pattern of loadings of $X_2$ and $X_4$ (they have low loadings on Factor 2), to which it has a link in Figure \ref{multstructure} (it is linked to $X_4$), than to the pattern of loadings of $X_5$ and $X_6$, to which it has no link. Similarly, the pattern of loadings for $X_7$ (high on Factor 2 and low on Factors 1 and 3) has more in common with the pattern of loadings of $X_5$ and $X_6$ (they have low loadings on Factor 3), to which it has a link, than to the pattern of loadings of $X_2$ and $X_3$, to which it has no link.
\end{enumerate}

Clearly the patterns of loadings in Table \ref{loadings} relate well to the links between variables in  Figure \ref{multstructure}. In some senses, Table \ref{loadings} and Figure \ref{multstructure} could not relate better. For example, if variables that are linked have different patterns of loadings then, for the structure in Figure \ref{multstructure}, at most two pairs of linked variables could have the same patterns of loadings, and that is the number of pairs with the same patterns in Table \ref{loadings} (c.f. Point 2). Hence it seems that the structure shown in Figure \ref{multstructure} is a meaningful reflection of real structure in the data.

\subsubsection*{3.4 Example 4: Artificial Data (two overlapping collinearities)} \label{S:Ex4}

A dataset of 100 observations on eight variables was constructed that had two collinearities. Data for $X_1$, $X_2$, $X_3$, $X_5$, $X_6$ and $X_7$ were independent values from a standard normal distribution, $X_4$ was a linear function of the first three variables:
\begin{equation} \label{Eq8a}
X_4 =  X _1+X_2+X_3 +0.25\epsilon_1
\end{equation} 
where $\epsilon_1 \sim \mbox{N}(0,1)$ and the second collinearity was formed from $X_4$ and $X_7$ as:
\begin{equation} \label{Eq9a}
X_8 =  X _4-X_7 +0.25\epsilon_2
\end{equation} 
where $\epsilon_2 \sim \mbox{N}(0,1)$. The collinearities overlap as both contain $X_4$. Data were standardised so that each variable had a mean of 0 and a variance of 1.

\begin{linespread}{1.5}
\begin{table}
\caption{\footnotesize Correlation matrix for the artificial data}\label{corartif} \vspace{-0.05in}
\newcommand\Fontvi{\fontsize{9.0}{9.0}\selectfont}
\Fontvi
\begin{center}
\begin{tabular}{c@{\hskip 0.25cm} c@{\hskip 0.25cm} c@{\hskip 0.25cm} c@{\hskip 0.25cm}  c@{\hskip 0.25cm} c@{\hskip 0.25cm} c@{\hskip 0.25cm} c@{\hskip 0.25cm} c}
\hline \vspace{-0.2in} \\ 
& $X_1$ & $X_2$ & $X_3$ & $X_4$ & $X_5$ & $X_6$ & $X_7$ & $X_8$ \\
\hline \vspace{-0.15in} \\
$X_1$ & \ 1.000 & \ 0.136 & \ 0.029 & \ 0.635 & \ 0.035 & \ 0.106 & \ 0.121 & \ 0.388 \\
$X_2$ & \ 0.136 & \ 1.000 & -0.018 & \ 0.608 & \ 0.009 & -0.181 & \ 0.022 & \ 0.461 \\
$X_3$ & \ 0.029 & -0.018 & \ 1.000 & \ 0.554 & \ 0.091 & -0.038 & \  0.045 & \  0.376 \\
$X_4$ & \ 0.635 & \ 0.608 & \ 0.554 & \ 1.000 & \ 0.066 & -0.085 & \  0.112 & \ 0.677 \\
$X_5$ & \ 0.035 & \ 0.009 & \ 0.091 & \ 0.066 & \ 1.000 & \ 0.271 & -0.078 & \ 0.121 \\
$X_6$ & \ 0.106 & -0.181 & -0.038 & -0.085 & \ 0.271 & \ 1.000 & \ 0.016 & -0.091 \\ 
$X_7$ & \ 0.121 & \ 0.022 & \ 0.045 & \ 0.112 &-0.078 & \ 0.016 & \ 1.000 & -0.632 \\
$X_8$ & \ 0.388 & \ 0.461 & \ 0.376 & \ 0.677 & \ 0.121 & -0.091 & -0.632 & \ 1.000 \\
				
\hline
\end{tabular}
\end{center}
\end{table}
\end{linespread}

Table \ref{corartif} presents the correlation matrix of the eight variables. Its eigenvalues are 2.713, 1.412, 1.331, 1.031, 0.862, 0.619,  \textcolor{blue}{0.022} and  \textcolor{blue}{0.008}. As there are two collinearities, there are two small eigenvalues and Table \ref{eigenartif} displays the corresponding eigenvectors along with the VIFs.  

In an eigenvector analysis, suppose 0.26 is taken as the cut-off for deciding which variables are included in a collinearity: a variable is included if and only the absolute value of its loading exceeds 0.26. Then $\boldsymbol{v}_{7}$ identifies a collinearity between all the variables except  $X_4$ $X_5$ and $X_6$, while  $\boldsymbol{v}_{8}$ identifies a collinearity between $X_1$, $X_2$, $X_3$ and $X_4$. Although not as simple as equations (\ref{Eq8a}) and (\ref{Eq9a}), these collinearities correctly identify the collinearities in those equations. However, incorrect collinearities are identified if a cut-off is used that is below 0.25 or above 0.3. With 0.3 as the cut-off, $\boldsymbol{v}_{7}$ incorrectly identifies a collinearity that involves just $X_7$ and $X_8$, while if the cut-off is taken as 0.25, then $\boldsymbol{v}_{8}$ incorrectly includes $X_8$ in the collinearity given by $X_1$, $X_2$, $X_4$ and $X_4$. With either 0.25, 0.26 or 0.3 used as the cut-off, the variables selected for the two collinearities are precisely those with large VIFs, so examining VIFS does not help select a cut-off. Consequently, if the mechanism that generated the data were unknown, the best cut-off would not be clear, and an eigenvector analysis is unlikely to identify collinearities that agree with equations (\ref{Eq8a}) and (\ref{Eq9a}).


\begin{linespread}{1.5}
\begin{table}
\caption{\footnotesize Eigenvector loadings and VIFs for the artificial data}\label{eigenartif} \vspace{-0.05 in}
\newcommand\Fontvi{\fontsize{9.0}{9.0}\selectfont}
\Fontvi
\begin{center}
\begin{tabular}{l@{\hskip 0.15cm} c@{\hskip 0.15cm} c@{\hskip 0.15cm} c@{\hskip 0.15cm} c@{\hskip 0.15cm} c@{\hskip 0.15cm} c@{\hskip 0.15cm} c@{\hskip 0.15cm} c }
\hline \vspace{-0.15in} \\ 
Eigenvector & $X_1$ & $X_2$ & $X_3$ & $X_4$ & $X_5$ & $X_6$ & $X_7$ & $X_8$ \\
\hline \vspace{-0.15in} \\
$\boldsymbol{v}_{7}$& \textcolor{blue}{0.289} & \ \textcolor{blue}{0.298} & \ \textcolor{blue}{0.283} & \ 0.007 & \ 0.013 & -0.025 & \textcolor{blue}{-0.507} & \textcolor{blue}{-0.700} \\
$\boldsymbol{v}_{8}$& \textcolor{blue}{-0.321} & \textcolor{blue}{-0.312} & \textcolor{blue}{-0.323} & \ \textcolor{blue}{0.772} & \ 0.006 & \ 0.009 & -0.187 & -0.253 \\
\hline \vspace{-0.15 in} \\
VIF & 16.85 & 16.35 & 16.77 & 70.77 & \ 1.12 & \ 1.19 & 16.35 & 30.06 \\
\hline
\vspace{-.2in}
\end{tabular}
\end{center}
\end{table}
\end{linespread}

Table \ref{vdpartif} presents results from the variance decomposition method for the two largest condition indices. The VDPs for the largest condition index correctly identify the collinearity between $X_1$, $X_2$, $X_3$ and $X_4$, based on the recommended threshold of 0.5. However, the VDPs for the second largest condition index incorrectly indicate a collinearity between $X_7$ and $X_8$.
Auxiliary regressions also give a mixed performance. The top row of Table \ref{vdpartif} indicates that either $X_7$ or $X_8$ should be the dependent variable in one regression. With either choice, the collinearity between $X_4$, $X_7$ and $X_8$ is correctly identified. The variable $X_4$ has much the largest VDP for $\boldsymbol{v}_{8}$ and the smallest VDP for $\boldsymbol{v}_{7}$, so it should be the dependent variable in the other auxiliary regression. In  a stepwise regression with $X_4$ as the dependent variable, variables are entered into the regression equation in the order $X_8$ (0.454), $X_{7}$ (0.943), $X_1$ (0.945), $X_3$ (0.952), $X_2$ (0.985). The $R^2$ values (in brackets) only increase slowly after $X_8$ and $X_{7}$ have been entered into the regression, so the analysis suggests a collinearity between $X_4$,  $X_8$ and $X_{7}$. However, this is the collinearity identified by the first auxiliary regression, so the correct inference might instead be made that there is collinearity between  $X_4$, $X_8$, $X_7$, $X_1$, $X_3$  and $X_2$.

\begin{linespread}{1.5} 
\begin{table}
\caption{Condition indices and variance-decomposition proportions for the artificial data}\label{vdpartif}\vspace{-0.05in}  
\newcommand\Fontvi{\fontsize{9.0}{9.0}\selectfont}
\Fontvi
\begin{center}
\begin{tabular}{ c@{\hskip 0.2cm} c@{\hskip 0.2cm} c@{\hskip 0.2cm} c@{\hskip 0.2cm} c@{\hskip 0.2cm}  c@{\hskip 0.2cm} c@{\hskip 0.2cm} c@{\hskip 0.2cm} c@{\hskip 0.2cm} c c }
\hline \vspace{-0.2in} \\ 
Eigen- & Condition & \multicolumn{8}{c}{Variance-decomposition proportion} \\
\cline{3-10}
vector & Index & $X_1$ & $X_2$ & $X_3$ & $X_4$ & $X_5$ & $X_6$ & $X_7$ & $X_8$ \\
\hline \vspace{-0.15in} \\
$\boldsymbol{v}_{7}$  & 123.3 & 0.225 & 0.246 & 0.218 & 0.000 & 0.007 & 0.024 & \textcolor{blue}{0.714} & \textcolor{blue}{0.740} \\
$\boldsymbol{v}_{8}$ & 321.5 &  \textcolor{blue}{0.727} & \textcolor{blue}{0.703} & \textcolor{blue}{0.738} & \textcolor{blue}{0.997} & 0.004 & 0.009 & 0.254 & 0.252  \\
\hline 
\end{tabular}
\end{center}
\end{table}
\end{linespread}

Table \ref{maxartif} shows the transformation matrix given by the cos-max method, together with VIFs. The VIFs equal $\ba_i^T \ba_i$ and they clearly show that all variables except for $X_5$ and $X_6$ are involved in collinearities -- the VIFs of $X_5$ and $X_6$ are well below 5 while other VIFs are well above 5. The transformation matrix also clearly identifies the collinearity in equation (\ref{Eq8a}) -- the elements of $\ba_1$ corresponding to $X_1$, $X_2$, $X_3$ and $X_4$ are all well above the threshold of 0.75 while other elements are comfortably below it, and the same is true of $\ba_2$ and  $\ba_3$. Similarly, the transformation matrix clearly identifies the collinearity in equation (\ref{Eq9a}) -- the elements of $\ba_7$ corresponding to $X_4$, $X_7$ and $X_8$ are well above 0.75 while other elements are comfortably below it, and the same is true of $\ba_8$. Since $X_4$ forms part of both collinearities, the elements of $\ba_4$ are large for all elements except $X_5$ and $X_6$. This example illustrates that, even when alternative methods struggle, the cos-max method may correctly identify overlapping collinearities with good clarity, repeating the message about a collinearity once for each variable involved in the collinearity.

\begin{linespread}{1.5} 
\begin{table}
\caption{\footnotesize Cos-max transformation matrix and VIFs for the artificial data}\label{maxartif} \vspace{-0.05in} 
\newcommand\Fontvi{\fontsize{9.0}{9.0}\selectfont}
\Fontvi
\begin{center}
\begin{tabular}{ l@{\hskip 0.15cm} c@{\hskip 0.15cm} c@{\hskip 0.15cm} c@{\hskip 0.15cm} c@{\hskip 0.15cm} c@{\hskip 0.15cm} c@{\hskip 0.15cm} c@{\hskip 0.15cm} c@{\hskip 0.15cm} c}
\hline \vspace{-0.15in} \\ 
& $X_1$ & $X_2$ & $X_3$ & $X_4$ & $X_5$ & $X_6$ & $X_7$ & $X_8$ & VIF  \\
\hline \vspace{-0.15in} \\
$\boldsymbol{a}_1^{\top}$ &  \ \textcolor{blue}{2.486} & \ \textcolor{blue}{1.372} & \ \textcolor{blue}{1.439} & \textcolor{blue}{-2.540} & \ 0.012 & -0.147 & -0.273 & -0.409 & 16.85 \\
$\boldsymbol{a}_{2}^{\top}$ & \ \textcolor{blue}{1.372} & \ \textcolor{blue}{2.459} & \ \textcolor{blue}{1.454} & \textcolor{blue}{-2.449} & -0.006 & \ 0.024 & -0.273 & -0.492 & 16.35 \\
$\boldsymbol{a}_{3}^{\top}$ & \ \textcolor{blue}{1.439} & \ \textcolor{blue}{1.454} & \ \textcolor{blue}{2.449} &  \textcolor{blue}{-2.527} & -0.037 & -0.047 & -0.223 & -0.387 & 16.77 \\
$\boldsymbol{a}_{4}^{\top}$ & \textcolor{blue}{-2.540} & \textcolor{blue}{-2.449} & \textcolor{blue}{-2.527} & \ \textcolor{blue}{6.750} & \ 0.029 & \ 0.094 & \textcolor{blue}{-1.455} & \textcolor{blue}{-2.061} & 70.77 \\
$\boldsymbol{a}_{5}^{\top}$ & \ 0.012 & -0.006 & -0.037 & \ 0.029 & \ 1.041 & -0.153 & -0.016 & -0.103 & \ 1.12 \\
$\boldsymbol{a}_{6}^{\top}$ & -0.147 & \ 0.024 & -0.047 & \ 0.094 & -0.153 & \ 1.059 & \ 0.046 & \ 0.100 & \ 1.19 \\
$\boldsymbol{a}_{7}^{\top}$ & -0.273 & -0.273 & -0.223 & \textcolor{blue}{-1.455} & -0.016 & \ 0.046 & \ \textcolor{blue}{2.717} & \ \textcolor{blue}{2.579} & 16.35 \\
$\boldsymbol{a}_{8}^{\top}$ & -0.409 & -0.492 & -0.387 & \textcolor{blue}{-2.061} & -0.103 & \ 0.100 & \ \textcolor{blue}{2.579} & \ \textcolor{blue}{4.311} & 30.06\\
\hline 
\end{tabular}
\end{center}
\end{table}
\end{linespread}
 
\subsection*{4. Concluding comments}\label{multconc}  

Three methods for identifying collinearities have been examined, eigenvector analysis, the variance decomposition method and the cos-max method. They approach the task of identifying collinearities from different directions. Eigenvector analysis and the variance decomposition method look at the eigenvector loadings for small eigenvalues, while the cos-max method examines how variables must be adjusted to make them orthogonal. Since they give different perspectives on a dataset, they may be treated as complementary methods rather than alternatives, as more than one method may be implemented to learn about the collinearities in a dataset.

The following are three attractive characteristics of the cos-max method that hold in general.
\begin{enumerate}
\item[(i)] {\em The cos-max method gives a coherent link between identifying which variables are involved in collinearities (VIFs are the standard criteria) and which variables are involved in each collinearity.} Consequently, in each of the four examples, the cos-max method gives collinearities that only contain variables with high VIFs (VIFs above 5) and together they contain all the variables with high VIFs.  In contrast, in Examples 2 and 3  the other methods make errors of inclusion (including a variable in a collinearity that has a low VIF) or omission (omitting a variable with a high VIF from all the collinearities it identifies). Errors of omission are made by the eigenvector analysis in Example 3, by the variance decomposition method in Examples 2 and 3, and by auxiliary regressions in Example 3. Errors of inclusion are made by the variance decomposition method in Examples 2 and 3, and by auxiliary regressions in Example 2.
\item[(ii)] {\em The cos-max method identifies collinearities using more information than alternative methods.} This is illustrated in each example. Eigenvector analysis and the variance decomposition method examine 1, 3, 3 and 2 eigenvectors in Examples 1, 2, 3 and 4, respectively, while the  cos-max method examines 4, 7, 8 and 6 $\ba$-vectors. Even when there is just a single collinearity and it is comparatively easy to identify, it is reassuring to see the identification confirmed in each of a number of rows of a table.
\item[(iii)]  {\em The cos-max method does not focus solely on collinearities and can identify other features in a data structure.} When there are $k$ small eigenvalues from the correlation matrix of $m$ regressors, it means that most of the variation in the regressors is contained in a hyperspace of dimension $m-k$. It does not mean that the relationships between the regressors can neatly be explained by $k$ linear relationships. It could, for example, arise because $m-k$ factors underlie the values taken by the $m$ regressors. The benefit from looking beyond collinearities is illustrated in Example 3 (c.f. Fig. 1).
\end{enumerate}   

When there is a single collinearity, in general the different methods will all identify it correctly (c.f Example 1).  When there are multiple collinearities the situation is more complex and inferences given by the different methods may vary. The following are some features of the cos-max method that emerged from the examples.
\begin{enumerate}
\item[(a)] {\em In all examples the collinearities proposed by the cos-max method were highly plausible.} As noted in point (i), the variables involved in the collinearities it proposed matched the variables with large VIFs. In Example 1 and 2 there was reasonable agreement between all methods in the collinearities they proposed and in Example 3 the cos-max method proposed a data structure that was consistent with the results of a factor analysis. Example 4 involves artificial data so the collinearities underlying the data structure are known. These collinearities were correctly identified by the cos-max method. 
\item[(b)]  {\em The cos-max method typically gives a set of collinearities that are as simple or simpler (more parsimonious) than those given by other methods.} For instance, in Example 2 it identified two disjoint collinearities, $\{X_1,X_2\}$ and $\{X_3,X_4\}$, while eigenvector analysis and the variance decomposition method identified two completely overlapping collinearities, each containing $X_1$, $X_2$, $X_3$ and $X_4$. In Example 4 the cos-max method identified two collinearities that only had one variable in common, while eigenvector analysis and auxiliary regressions identified collinearities that had three variables in common. Also, the cos-max transformation can suggest partial correlations that will be informative and help it form parsimonious collinearities, as in Example 3.

\end{enumerate}

In summary, the cos-max method is a useful addition to a statistician's toolbox as a means of identifying collinearities and exploring the structure of a dataset.

\subsection*{Appendix: Inference from the transformation matrix} \label{S:Append}

This appendix addresses the question of which elements of the transformation matrix ($\bA$) will be large and which will be small.
 
The purpose of the transformation is to find a set of orthogonal vectors of unit length, $\bu_1, \ldots, \bu_m$, such that $\psi =\sum_{i=1}^m \bx_i^T \bu_i $ is as large as possible (c.f. equation (\ref{eqPsi}). Now $\bx_i^T \bu_i = \cos \theta_i$ where $\theta_i$ is the angle between the vector from $\bzero$ to $\bx_i$ and the vector from $\bzero$ to $\bu_i$. Thus $\bx_i^T \bu_i$ has a maximum value of 1 when $\theta_i =0$ and its value decreases as $\theta$ increases. Hence the transformation aims to minimise the differences between the $\bx$-vectors and their surrogates. If  $\bu_i$ differs little from $\bx_i$, then the $i$th element of $\ba_i$ will be close to 1 and other elements of $\ba_i$ will be small. Except where correlations between $X$ variables prevent it, off-diagonal elements of the transformation matrix $\bA$ will be small.

When the variables $X_i$ and $X_j$ are correlated, $\bx_i^T \bx_j \neq 0$. Since $\bu_i^T \bu_j = 0$, it follows that $\bx_i$ and $\bx_j$ cannot both equal their surrogates. The matrix $\bA$ is symmetric, so the proportion of $\bx_i$ that contributes to $\bu_j$ is equal to the proportion of $\bx_j$ that contributes to $\bu_i$. Hence, typically both $\bx_i$ and $\bx_j$ will differ from their surrogates when $X_i$ and $X_j$ are correlated. If $X_i$ and $X_j$ are uncorrelated with the other $X$ variables, then only the $i$th and $j$th elements of $\ba_i^T$ and $\ba_j^T$ will be non-zero:
\begin{equation} \label{EqNonZero}
\bu_i=a_{ii}\bx_i +a_{ij}\bx_j;  \quad \quad \quad \bu_j=a_{ji}\bx_i +a_{jj}\bx_j 
\end{equation}
and $a_{ji} =a_{ij}$. Table (\ref{corrvec}) gives the values of $a_{ii}$ and $a_{ij}$ for various correlations between $X_i$ and $X_j$ (with no correlation between them and other variables). From the table, the VIF exceeds a threshold of 5 when the correlation is about 0.9, if the variables are not correlated with other variables. We have taken off-diagonal elements of $\bA$ above 0.75 as indicative of which variables contribute to a collinearity.

\begin{linespread}{1.5} 
\begin{table}
\caption{Values of $a_{ii}$ and $a_{ij}$ for different correlations between $X_i$ and $X_j$ when they are uncorrelated with other $X$ variables} \label{corrvec} \vspace{-0.02in}
\newcommand\Fontvi{\fontsize{10.0}{10.0}\selectfont}
\Fontvi
\begin{center}
\begin{tabular}{ l@{\hskip 0.35cm} c@{\hskip 0.35cm} c@{\hskip 0.35cm} c@{\hskip 0.35cm} c@{\hskip 0.35cm} c@{\hskip 0.35cm} c@{\hskip 0.35cm} c@{\hskip 0.35cm} c }
\hline \vspace{-0.2in} \\ 
Correlation& 0 \  & 0.2 & 0.4 & 0.6 & 0.8 & 0.9 & 0.95 & 0.99 \\
\hline \vspace{-0.15in} \\
$a_{ii}$ & 1 \ & \ 1.015 & \ 1.068 & \ 1.186 & \ 1.491 & \ 1.944 & \ 2.594 & \ 5.354 \\ 
$a_{ij}$ & 0 \ & -0.103 & -0.223 & -0.395 & -0.745 & -1.218 & -2.878 & -4.646 \\
\hline \vspace{-.4in}
\end{tabular}
\end{center}
\end{table}
\end{linespread} 

The value of $a_{ij}$ is influenced not just by the correlation between $X_i$ and $X_j$, but also by their correlations with other $X$ variables. In particular, correlations that result in a collinearity will have a marked effect on the elements of $\bA$. This is because a collinearity between, say, $k$ variables implies that the values of these variables lie close to a hyperplane whose dimension is less than $k$. In contrast, the surrogates of these $X$-variables form a basis for a $k$-dimensional hyperplane. Thus there must be substantial differences between the $X$-variables from a collinearity and their surrogates and the corresponding elements of $\bA$ will be large. 

The matrix $\bA$ can be related to a spectral decomposition of $\bX^T\bX $. Let $\lambda_1 \geq \lambda_2 \geq \ldots \geq \lambda_m$ denote the eigenvalues of $\bX^T \bX$ and let $\bv_1, \ldots , \bv_m$ be the corresponding eigenvectors. 
 From the spectral decomposition theorem,
\begin{equation} \label{SpecEq}
\bX^T\bX =(\bv_1, \ldots, \bv_m)
\left( \begin{array}{ccc}
\lambda_1 & &\bzero\\
 & \ddots & \\
\bzero & & \lambda_m
\end{array} \right)
\left( \begin{array}{c}
\bv_1^T\\
 \vdots \\
\bv_m^T
\end{array} \right)
\end{equation}
so, since the transformation matrix $\bA$ equals $(\bX^T \bX)^{-1/2}$,
\begin{equation} \label{SpecA}
\bA =(\bv_1, \ldots, \bv_m)
\left( \begin{array}{ccc}
\lambda_1^{-1/2} & &\bzero\\
 & \ddots & \\
\bzero & & \lambda_m ^{-1/2}
\end{array} \right)
\left( \begin{array}{c}
\bv_1^T\\
 \vdots \\
\bv_m^T
\end{array} \right) .
\end{equation}
If there is a single collinearity and that collinearity is strong, then $\lambda_m$ will be much smaller than the other $\lambda_i$, so $\lambda_m^{-1/2}$ will be much larger than the other $\lambda_i^{-1/2}$. Also, the elements of $\bv_m$ that differ markedly from 0 will correspond to the $X$ variables that form the collinearity, while other elements of $\bv_m$ will be close to 0. If a vector $\bv_{m\#}$ is formed by setting to 0 the elements of $\bv_m$ that are small in absolute value (so $\bv_{m\#} \approx \bv_m$), then 
\begin{equation} \label{Eq12}
\bA= \bv_{m\#}\lambda_m^{-1/2}\bv_{m\#}^T +\bB
\end{equation}
where all elements of $\bB$ are small relative to every non-zero element of  $\bv_{m\#}\lambda_m^{-1/2}\bv_{m\#}^T$. Thus the large elements of $\bA$ identify the collinearity as they correspond to the non-zero elements of $\bv_{m\#}\bv_{m\#}^T$. This is illustrated in Example 1 (c.f $\bv_5$ in Table \ref{eigensales} with Table \ref{garsales1}).

The situation is more complicated when there is more than one collinearity. If $\lambda_m$ is much smaller than the other eigenvalues then, as with the case of one strong collinearity, the non-zero elements of $\bv_{m\#}\bv_{m\#}^T$ will cause the corresponding elements of $\bA$ to be large. For other eigenvalues the situation is less clear, as commonly an eigenvector associated with a less extreme eigenvalue will not dominate the structure of $\bA$. This is illustrated in Examples 2, 3 and 4, where the eigenvectors corresponding to small eigenvalues do not identify the same collinearities as the cos-max method.

The transformation matrix will contain many zeros if the $X$ variables can be divided into sets so that variables in different sets are uncorrelated with each other. To simplify notation suppose, for definiteness, that there are three sets. Order the variables so that $X_1, \ldots, X_k$ are in the first set, $X_{k+1}, \ldots, X_l$ are in the second set, and $X_{l+1}, \ldots, X_m$ are in the third set. Then, for example, every variable from $X_{k+1}, \ldots, X_l$ is uncorrelated with every element from $X_1, \ldots, X_k$ and every element from $X_{l+1}, \ldots, X_m$. Hence the matrix $\bX^T \bX$ has the block-diagonal form
\begin{equation}
\bX^T \bX = \left( \begin{array}{ccc}
\bR_1 & \bzero & \bzero \\
\bzero & \bR_2 & \bzero \\
\bzero & \bzero & \bR_3
\end{array}
\right),
\end{equation}
where $\bR_i$ is the correlation matrix of the $i$th set of variables ($i=1,2,3$). Then the transformation matrix is
\begin{equation}
\bA = (\bX^T \bX)^{-1/2} = \left( \begin{array}{ccc}
\bR_1^{-1/2} & \bzero & \bzero \\
\bzero & \bR_2^{-1/2} & \bzero \\
\bzero & \bzero & \bR_3^{-1/2}
\end{array}
\right).
\end{equation}
For $i=1,\ldots,k$, the first $k$ elements of $\ba_i$ are determined by $\bR_1$ (the correlation matrix of $X_1, \ldots, X_k$) and the remaining components of $\ba_i$ are 0; also $\bu_i = \bX \ba_i$ is determined solely by $\bR_1$. Similar comments apply to $\ba_{k+1}, \ldots, \ba_l$ and $\ba_{l+1}, \ldots, \ba_m$. 
Returning to the general case, if the $X$ variables can be partitioned into sets so that variables in different sets are uncorrelated, then the $i$th component of $\ba_j$ will be 0 if $X_i$ and $X_j$ are in different sets. Extending the argument,  if the $X$ variables can be partitioned into sets so that variables in different sets have low correlation, then the $i$th component of $\ba_j$ will be small if $X_i$ and $X_j$ are in different sets. One consequence relates to datasets in which there are two or more disjoint collinearities. Suppose variables are partitioned into sets with low correlation between the sets. If each set contains variables from at most one collinearity, then examination of the transformation matrix will separate the different collinearities; the large values in $\bA$ will not suggest constructing an unnecessarily complex collinearity by combining disjoint collinearities.

\subsubsection*{References}\vspace{-.02in}
\begin{enumerate}
\item [[1]] D.A. Belsley, {\em Conditioning Diagnostics: Collinearity and Weak Data in Regression}, John Wiley and Sons, New York, 1991.\vspace{-.12in} 
\item[[2]] D.A. Belsley, E. Kuh, and R.E. Welsch, {\em Regression Diagnostics: Identifying Influential Data and Sources of Collinearity}, John Wiley and Sons, New York, 1980.\vspace{-.12in} 
\item[[3]] S. Chatterjee, and A.S. Hadi, {\em Regression Analysis by Example}, John Wiley and Sons, New York, 2006.\vspace{-.12in} 
\item[[4]] P.H. Garthwaite, F. Critchley, K. Anaya-Izquierdo, and E. Mubwandarikwa, {\em Orthogonalization of vectors with minimal adjustment}, Biometrika 99 (2012), pp.787--798.\vspace{-.12in}
\item[[5]] D.N. Gujarati, {\em Basic Econometrics}, McGraw-Hill, Boston, 2003.\vspace{-.12in}
\item[[6]] R.R. Hocking, {\em Methods and Applications of Linear Models: Regression and the Analysis of Variance}, John Wiley and Sons, New York, 2003.\vspace{-.12in} 
\item[[7]] J.N.R. Jeffers, {\em Two case studies in the application of principal component analysis}, J. R. Stat. Soc. Ser. C Appl. Stat. 16 (1967), pp.225--236.\vspace{-.12in}
\item[[8]] V. Mahajan, A.K. Jain, and M. Bergier, {\em Parameter estimation in marketing models in the presence of multicollinearity: an application of ridge regression}, J. Mark. Res. 14 (1977), pp.586--591.\vspace{-.12in}
\item[[9]] D.C. Montgomery, E.A. Peck, and G.G. Vining, {\em Introduction to Linear Regression Analysis}, John Wiley and Sons, New York, 2015.\vspace{-.12in}
\item[[10]] J. Neter, W. Wasserman, and M.H. Kutner, {\em Applied Linear Regression Models,} Irwin, Illinois, 1983.\vspace{-.12in}
\item[[11]] C. Ofir and A. Khuri, {\em Multicollinearity in marketing models: diagnostics and remedial measures}, Int. J. Res. Mark. 3 (1986), pp.181--205.\vspace{-.12in}
\end{enumerate}

\end{document}